\documentclass[prl, twocolumn, preprintnumbers,amsmath,amssymb,superscriptaddress,longbibliography]{revtex4-2}

\usepackage{graphicx}% Include figure files
\usepackage{dcolumn}% Align table columns on decimal point
\usepackage{bm}% bold math
\usepackage{subfigure}
\usepackage{multirow}
\usepackage{amsmath}
\usepackage{color,soul}
\usepackage[font=footnotesize, justification=raggedright,singlelinecheck=false]{caption}
\usepackage{soul}
\usepackage{cancel}

\usepackage[colorlinks=true, linkcolor=blue, urlcolor=blue, citecolor=blue]{hyperref}
\newcommand{\add}[1]{{\color{black} #1}}

\begin{document}
\title{Liquid-Gas Criticality of Hyperuniform Fluids}

\author{Shang Gao}
\thanks{These authors contributed equally.}
\author{Hao Shang}
\thanks{These authors contributed equally.}
\affiliation{National Laboratory of Solid State Microstructures,  Nanjing University,  Nanjing 210093, China}
\affiliation{Collaborative Innovation Center of Advanced Microstructures and School of Physics,  Nanjing University,  Nanjing 210093, China}
\affiliation{Jiangsu Physical Science Research Center, Nanjing 210093, China}

\author{Hao Hu}
\affiliation{School of Physics, Anhui University, Hefei 230601, China}

\author{Yu-Qiang Ma}
\email{myqiang@nju.edu.cn}
\affiliation{National Laboratory of Solid State Microstructures,  Nanjing University,  Nanjing 210093, China}
\affiliation{Collaborative Innovation Center of Advanced Microstructures and School of Physics,  Nanjing University,  Nanjing 210093, China}
\affiliation{Jiangsu Physical Science Research Center, Nanjing 210093, China}
\affiliation{Hefei National Laboratory, Shanghai 201315, China}

\author{Qun-Li Lei}
\email{lql@nju.edu.cn}
\affiliation{National Laboratory of Solid State Microstructures,  Nanjing University,  Nanjing 210093, China}
\affiliation{Collaborative Innovation Center of Advanced Microstructures and School of Physics,  Nanjing University,  Nanjing 210093, China}
\affiliation{Jiangsu Physical Science Research Center, Nanjing 210093, China}
\affiliation{Hefei National Laboratory, Shanghai 201315, China}

\begin{abstract}
In statistical physics, it is well established that the liquid-gas (LG) phase transition with divergent critical fluctuations belongs to the Ising universality class. Whether non-equilibrium effects can alter this universal behavior remains a fundamental open question. In this work,  we theoretically prove that non-equilibrium hyperuniform (HU) fluids with additional center-of-mass conservation exhibit LG criticality different from the Ising universality class. As a specific case, we investigate a 2D HU fluid composed of active spinners, where phase separation is driven by dissipative collisions. Strikingly, at the critical point, the 2D HU fluid displays finite density fluctuations $S(q)\sim q^{\eta} \sim const.$  with $\eta=0$, rather than the expected divergence $S(q)\sim q^{\eta-2}$ with $\eta=1/4$ as in the Ising model, while the compressibility still diverges. The critical point is thus calm yet highly susceptible, in fundamental violation of the conventional fluctuation-dissipation relation. Consistently, we observe short-range \add{pair} correlation functions coexisting with quasi-long-range response functions at the critical point. Based on a generalized Model B and renormalization-group analysis, we prove that hyperuniformity reduces the upper critical dimension $d_c$ from $4$ to $2$. Moreover, the critical point exhibits Gaussian density fluctuations indicated by Binder cumulant, distinct from non-Gaussian critical behaviors of the mean-field Ising universality class. The system also exhibits non-divergent energy fluctuations rather than logarithmic divergence as expected in the Ising universality class at $d=d_c$. Furthermore, the HU fluid undergoes non-conventional spinodal decomposition, where the decomposition time diverges but the characteristic length scale remains finite as the critical point is approached. The origin of the above anomalies lies in the non-equilibrium nature of the system which obeys a generalized fluctuation-dissipation relation $2\mathrm{Im}~ \chi(q,\omega) ={\omega }C(q,\omega)/{k_B T_{\text{eff}}(q)}$ with a scale-dependent effective temperature $T_{\rm eff}(q) \propto q^2$. These findings establish a striking exception to conventional paradigms of critical phenomena and illustrate how non-equilibrium forces can fundamentally reshape universality classes.
\end{abstract}
\maketitle

\section{Introduction}

The liquid-gas (LG) phase transition is a fundamental thermodynamic phenomenon observed across a wide scale of matter~\cite{siemens1983liquid,lunine2008methane}. At its critical point, the distinction between liquid and gas phases disappears, giving rise to divergent density fluctuations and striking phenomena such as critical opalescence~\cite{chu1972critical}. 
{In the 1950s, Lee and Yang proved the equivalence between the LG phase transition in the lattice gas model and the order-disorder transition in the Ising model. After extensive theoretical~\cite{halperin1976renormalization,hohenberg1977theory,CriDyn}, numerical~\cite{Bruce1992,Watanabe2012,Yarmolinsky2017,
Wilding1992,Wilding1997} and experimental validation~\cite{Hayes1977,Pestak1984,Kim1984}, it is now well accepted that LG criticality of equilibrium fluids with short-range interactions lies in the Ising universality class~\cite{Lee1952,Yang1964}. }

For non-equilibrium systems, however, criticality need not follow the same universality as their equilibrium counterparts~\cite{tauber1999non,santos2002non,
speck2022critical,bertrand2022diversity,jentsch2023critical,miller2024phase}. 
Nonetheless, it has been increasingly recognized that an effective temperature and the corresponding Ising universality class can still emerge in non-equilibrium order-disorder transitions, LG transitions, or binary phase separation, provided no additional symmetry breaking or long-range interactions are present~\cite{Young2020,Han2017,nakano2021long,ikeda2024dynamical}. {In soft matter systems,} a prominent example is motility-induced phase separation (MIPS), which has been reported in active Brownian particle systems~\cite{Partridge2019,Dittrich2023,feng2025critical}, active Ornstein-Uhlenbeck particle systems~\cite{paoluzzi2016critical,Maggi2021,Maggi2022}, and quorum-sensing active particle systems~\cite{Gnan2022}. Other paradigmatic cases include the non-equilibrium kinetic Ising model~\cite{gambetta2019classical,di2025off,sides1998kinetic}, the two-temperature model~\cite{szolnoki1999stationary}, interacting run-and-tumble models~\cite{ray2024motility}, noise-induced phase separation~\cite{paoluzzi2024noise,paoluzzi2022scaling}, driven colloidal mixtures~\cite{Han2017} {and cellular
automata~\cite{takeuchi2006can}.}
These developments, along with ongoing debates~\cite{siebert2018critical,dittrich2021critical,Yarmolinsky2018,
saha2024site,bhowmick2025geometric}, raise the question of whether the Ising universality class is truly unique to LG criticality in fluids, regardless of whether the system is in or out of equilibrium.

In parallel, disordered hyperuniformity has emerged as a distinctive hallmark of certain disordered {systems}, characterized by strongly suppressed long-wavelength density fluctuations~\cite{Torquato2003,Tjhung2015,hexner2015hyperuniformity,PhysRevLett.118.020601,ma2017random,Torquato2018,Wilken2020,Chen2021,oppenheimer2022hyperuniformity,
lei2023does,zheng2024universal,chen2024emergent,
liu2024universal,wang2025active,
ma2025hyperuniformity}. Recently, the hyperuniform (HU) fluid state has been proposed and realized in active matter systems, including circle swimmers~\cite{Lei2019,Huang2021,Zhang2022}, active spinners~\cite{Lei2019h,oppenheimer2022hyperuniformity,
liu2023local,farhadi2018dynamics}, pulsating cells~\cite{li2025fluidization,keta2025long} \add{and vibrated granular systems~\cite{maire2024interplay}}. The hydrodynamic mechanism underlying HU fluids is that reciprocal active forces, combined with kinetic energy damping, result in effective center-of-mass conservation at large length scales~\cite{PhysRevLett.118.020601,Lei2019h}, which also induces other interesting phenomena~\cite{bertrand2019nonlinear,galliano2023two,ikeda2023correlated,
ikeda2024harmonic,kuroda2025long}. 
In quantum systems, similar center-of-mass/dipole conservation can stabilize exotic fractonic matter such as fracton fluids~\cite{gromov2020fracton,yuan2020fractonic,guardado2020subdiffusion, glorioso2022breakdown,han2024scaling,nandkishore2019fractons}.
These discoveries motivate a fundamental inquiry: Can the universality class of LG phase transitions be altered in HU fluids with center-of-mass conserved dynamics?

\begin{figure*}[!bhtp]
	\resizebox{180mm}{!}{\includegraphics[trim=0.0in 0.0in 0.0in 0.0in]{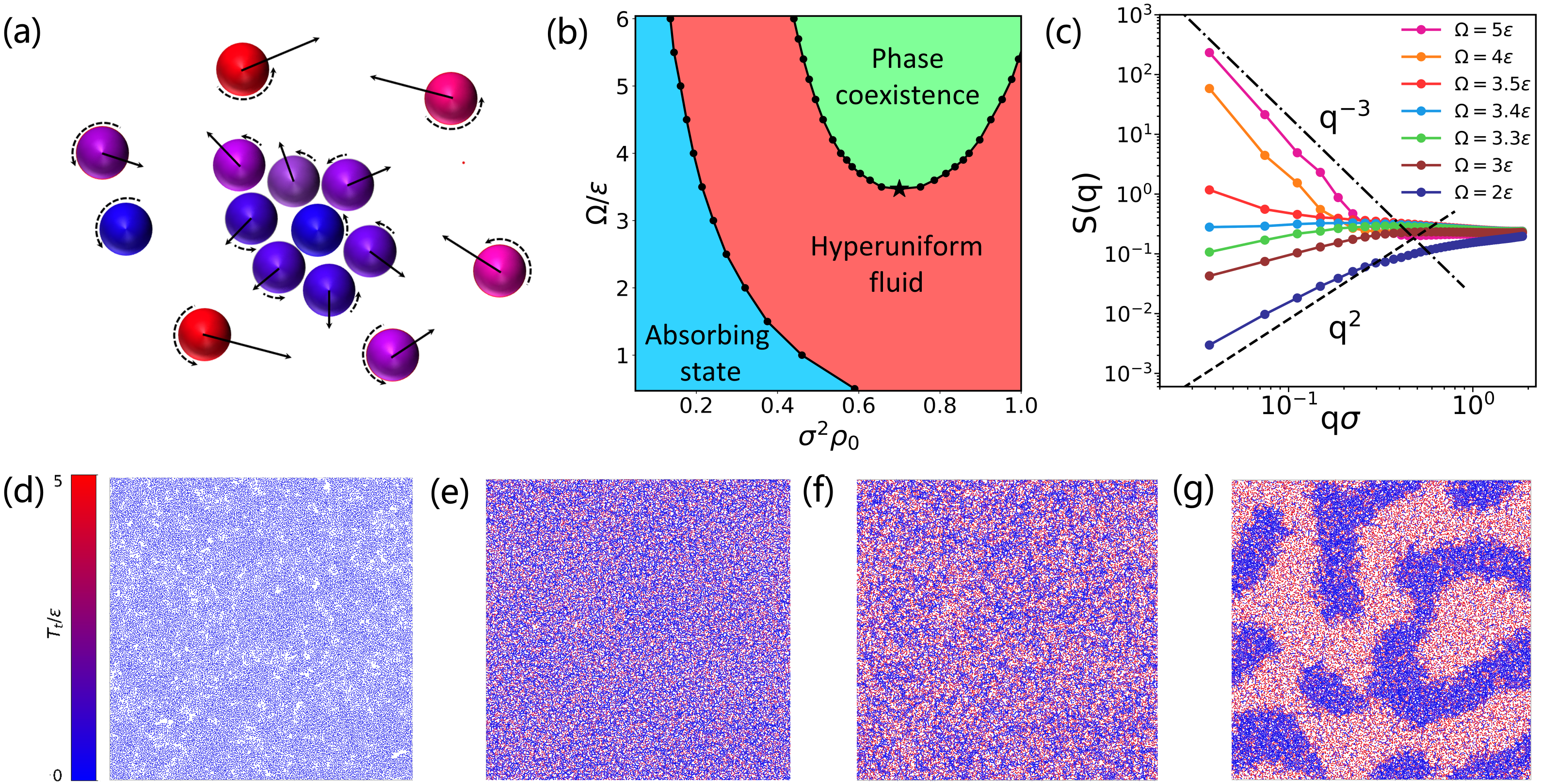} }
\caption{ (a) Schematic of dissipation-induced phase separation (DIPS) of {one-component} active spinner fluids, where spinner color (from blue to red) {encodes} the magnitude of spinners' translational kinetic energy $T_t$. (b) The phase diagram of system in dimensions of density $\rho$ and driven torque $\Omega$, where the absorbing state (cyan), HU fluids (red), phase coexistence (green) regimes are depicted. (c)  Structure factor $S(q)$ under different $\Omega$ at the critical density $\rho_c=0.7\sigma^{-2}$. Snapshots of absorbing state (d), HU fluids (e), LG critical point (f), phase {coexistence} (g). {Corresponding movies (Movie S1-S4)  are  provided in SM~\cite{supmat}}. }\label{MainFigSpinner}
\end{figure*}

In this work, based on the field theory and molecular dynamics simulations, we prove that the LG criticality of HU fluids is distinct from the Ising universality class. {As a specific case}, we investigate the LG criticality of a HU fluid composed of active spinners, where phase separation is induced by dissipative collisions. Strikingly, unlike conventional LG criticality characterized by divergent density fluctuations {$S(q) \sim q^{\eta-2}$ with ${\eta=1/4}$}, the {2D} HU fluid exhibits {finite} density fluctuations {$S(q)\sim q^{\eta}$ with ${\eta=0}$. The critical HU fluid also exhibits a short-range pair correlation instead of a quasi-long-range correlation.}
Based on {a} generalized Model B and renormalization-group analysis, we show that hyperuniformity reduces the upper critical dimension from $d_c = 4$ (conventional fluids) to $d_c = 2$. Stochastic field simulations confirm this reduction of $d_c$ and {further reveal a quasi-long-range response function in contrast with the short-range  correlation function at criticality. In addition, the theory and simulations also confirm the  divergent compressibility along with finite \add{density} fluctuations at criticality.} Thus, HU fluids are unexpectedly \emph{calm yet extremely susceptible} at the LG critical point, in fundamental violation of the conventional fluctuation-dissipation relation {(FDR). Moreover, the critical point exhibits Gaussian density fluctuations indicated by Binder cumulant, distinct from non-Gaussian critical behaviors of the mean-field Ising universality class. The system also exhibits non-divergent  energy fluctuations rather than logarithmic divergence as expected in the Ising universality  class at  $d=d_c$}. Finally, we observe non-conventional spinodal decomposition of HU fluids: as the critical point is approached, the decomposition time diverges while the characteristic length scale remains finite.

The origin of all these anomalies lies in the non-equilibrium dynamics of the system, which conserves the center-of-mass at large scales and effectively generates a scale-dependent temperature $T_{\rm eff}(q) \propto q^2$ that underlies a generalized FDR for HU fluids in Fourier space, i.e., $2\mathrm{Im}~ \chi(q,\omega) ={\omega }C(q,\omega)/{k_BT_{\text{eff}}(q)}$. Thus, the calm yet highly susceptible nature of HU fluids can be simply understood as $T_{\rm eff}$ approaching zero at long wavelengths.
These findings establish a striking exception to conventional paradigms of LG critical phenomena and demonstrate how non-equilibrium effects can fundamentally reshape both static and dynamic universality classes.

The paper is organized as follows:
in Sec.~II, we introduce the general field theory of HU fluids based on conservation law and prove that it satisfies a generalized FDR; 
in Sec.~III, we present a specific microscopic model of HU fluids composed of active spinners;
in Sec.~IV, we demonstrate that dissipative collisions between spinners induce liquid-gas phase separation, map out the corresponding phase diagram and reveal the violation of the conventional FDR;
in Sec.~V, we derive a hydrodynamic theory for the active spinner fluid and extract an effective Model B-like field theory that captures the criticality of phase separation; 
in Sec.~VI, we perform renormalization-group analysis on this effective theory and predict a reduction of the upper critical dimension along with non-conventional scaling behaviors at criticality; 
in Sec.~VII, we validate these predictions via large-scale stochastic field simulations; 
in Sec.~VIII, we investigate the spinodal decomposition and coarsening dynamics of HU fluids. We conclude in Sec.~IX with a discussion of future perspectives. Several appendixes along with Supplemental Materials (SM)~\cite{supmat} provide detailed derivations and \add{supplemental data} for interested readers (see also references ~\cite{thompson2022lammps,panagiotopoulos1994molecular,
  kubo1957statistical,newman1999monte,rossi2000universality,
  basu2012fixed,lee2013comment} therein).

\section{Non-equilibrium Field Theory of HU Fluids }

Previous theoretical works have established that HU fluids can be described by a minimal field theory with only a diffusion term and a center-of-mass  conserved noise term~\cite{PhysRevLett.118.020601,Lei2019h}. In this section, we proceed to develop the non-equilibrium field theory of phase separation in HU fluids. We consider systems described by a conserved order parameter, such as a density field $\psi(\mathbf{x},t)$. Its dynamics is described by continuity equations
\begin{eqnarray}
    \frac{\partial \psi(\mathbf{x},t)}{\partial t} &=& - \nabla\cdot \mathbf{J}(\mathbf{x},t)  \label{CMC-modelB-minimal1}  \\
     \mathbf{J}(\mathbf{x},t) &=&  -\nabla \left[  \add{\mathcal{J}[\psi](\mathbf{x},t)} + {\eta_{\psi}}(\mathbf{x},t) \right].   \label{CMC-modelB-minimal2}
\end{eqnarray}
Here, as the constraint of center-of-mass  conservation, the flux $\mathbf{J}(\mathbf{x},t)$ is written as the gradient of the field $\mathcal{J}[\psi](\mathbf{x},t) +{\eta_{\psi}}(\mathbf{x},t)$, where  {$\mathcal{J}[\psi](\mathbf{x},t)$ and ${\eta_{\psi}}(\mathbf{x},t)$} are the deterministic and noise terms, respectively. $\mathcal{J}[\psi](\mathbf{x},t)$ here is not necessarily  a derivative of a \add{effective} free energy functional. ${\eta_{\psi}}(\mathbf{x},t)$ is Gaussian white noise $\langle{ \eta_{\psi}(\mathbf{x},t)\eta_{\psi}(\mathbf{x}',t') }\rangle = 2D \delta^d(\mathbf{x}-\mathbf{x}')\delta(t-t')$, where $D$ is the noise strength. Note that Eq.~(\ref{CMC-modelB-minimal1}-\ref{CMC-modelB-minimal2}) can be viewed as a subclass of a general field theory with correlated noise~\cite{ikeda2023correlated}. 
Generally $\mathcal{J}[\psi](\mathbf{x},t)$ and $\eta_{\psi}(\mathbf{x},t)$ can also be tensor fields~\cite{de2024hyperuniformity}. In Appendix A, we prove that Eq.~(\ref{CMC-modelB-minimal1}-\ref{CMC-modelB-minimal2}) is an intrinsically non-equilibrium dynamic equation satisfying a generalized FDR in Fourier space
\begin{eqnarray}\label{GFDR}
    \mathrm{Im}~ \chi(q,\omega) &=& \frac{\omega }{2k_B T_{\text{eff}}(q)}C(q,\omega)
\end{eqnarray}
with  $k_BT_{\text{eff}}(q)$ the scale dependent effective temperature
\begin{eqnarray}\label{GFDR2}
    k_BT_{\text{eff}}(q) &=& Dq^2.
\end{eqnarray}
$\chi(q,\omega)$ and $C(q,\omega)$ \add{In Eq.~(\ref{GFDR})} are the dynamic response function and correlation function of fluids, respectively. Numerical verifications of this relation \add{will be provided later} in Fig.~\ref{MainFigGFDR}. Intuitively, \add{Eq.(\ref{GFDR}) indicates that} HU fluids are cooler at a larger length scale, indicating that HU fluids are non-equilibrium systems at a fundamental level. Only when $\mathcal{J}[\psi](\mathbf{x},t)$ is double space derivative of another field, i.e.,  $\mathcal{J}[\psi](\mathbf{x},t) =-\nabla^2 g[\psi](\mathbf{x},t)$, the conventional FDR can be recovered, and Eq.~(\ref{CMC-modelB-minimal1}-\ref{CMC-modelB-minimal2}) reduces to the field theory for an equilibrium system with center-of-mass  conservation~\cite{han2024scaling}. 
When $\mathcal{J}[\psi](\mathbf{x},t)$ becomes non-monotonic, a bifurcation would happen leading to multiple fixed points. Thus, Eq.~(\ref{CMC-modelB-minimal1}-\ref{CMC-modelB-minimal2}) provides a generalized theoretical framework to study the phase separation and corresponding critical phenomena in HU fluids. 
Due to the fundamental non-equilibrium nature characterized by Eq.~(\ref{GFDR}), the LG criticality of HU fluids might be different from the Ising universality class. \add{This difference is a focus} of this work.

\section{Model of Active Spinner Fluid}
As a specific case of the above theory, we study a system of $N$ disk-shaped spinners with mass $m$ and diameter $\sigma$ on a planar substrate (Fig.~\ref{MainFigSpinner}(a)). Each spinner is driven by a constant rotational torque $\boldsymbol{\Omega}$, experimentally achievable through various actuation methods including electric motors, magnetic fields or acoustic excitation  etc.~\cite{Zuiden2016,aragones2016elasticity,Kokot2017,ShieldsIV2018,Sabrina2018,Scholz2018a,farhadi2018dynamics,liu2020oscillating,yang2021topologically,Gardi2023,li2023inertial,modin2023hydrodynamic,li2024memory,wang2024robo,chen2024self}.

The dynamics of spinner $i$ is described by underdamped equations for both translational and rotational motion:
\begin{align}
    m\dot{\mathbf{v}}_i &= -\gamma_t \mathbf{v}_i + \sum_{j \neq i} \mathbf{f}_{ij}, \label{motionEqu} \\
    I\dot{\boldsymbol{\omega}}_i &= -\gamma_r \boldsymbol{\omega}_i + \sum_{j \neq i} \mathbf{r}_{ij} \times \mathbf{f}_{ij} + \boldsymbol{\Omega}, \label{rotationEqu}
\end{align}
where $\mathbf{v}_i = \dot{\mathbf{r}}_i$ and $\boldsymbol{\omega}_i$ represent translational and angular velocities, respectively; $I$ denotes the moment of inertia; $\gamma_t$ and $\gamma_r$ are the translational and rotational friction coefficients.
The pairwise interaction $\mathbf{f}_{ij}$ between spinners $i$ and $j$ incorporates three distinct components~\cite{brilliantov1996,shafer1996}:
\begin{equation}
    \mathbf{f}_{ij} = \mathbf{f}_{e,ij} + \mathbf{f}_{d,ij} + \mathbf{f}_{t,ij}, \label{force_components}
\end{equation}
where $\mathbf{f}_{e,ij}$ is the Hertzian elastic repulsion, $\mathbf{f}_{d,ij}$ is a velocity-dependent dissipative normal force associated with inelastic collision and $\mathbf{f}_{t,ij}$ is tangential friction:
\begin{align}
    \mathbf{f}_{e,ij} &= k_e (\sigma - r_{ij})^{3/2} \hat{\mathbf{r}}_{ij}, \label{elastic_force} \\
    \mathbf{f}_{d,ij} &= -k_d (\sigma - r_{ij})^{1/2} \left[ (\mathbf{v}_i - \mathbf{v}_j) \cdot \hat{\mathbf{r}}_{ij} \right] \hat{\mathbf{r}}_{ij}, \label{dissipative_force} \\
    \mathbf{f}_{t,ij} &= -k_t \mathbf{V}_{ij}. \label{tangential_force}
\end{align}
Here, $\hat{\mathbf{r}}_{ij} = \mathbf{r}_{ij}/r_{ij}$, $k_e$ and $k_d$ are elastic and dissipative coefficients, and $k_t$ is the tangential stiffness. 
The relative tangential velocity at contact point $\mathbf{V}_{ij}$ combines rotational and translational contributions: 
\begin{equation}
    \mathbf{V}_{ij} = \underbrace{\frac{\sigma}{2} (\boldsymbol{\omega}_i + \boldsymbol{\omega}_j) \times \hat{\mathbf{r}}_{ij}}_{\text{rotational slip}} + \underbrace{\left[ (\mathbf{v}_i - \mathbf{v}_j) \cdot \hat{\mathbf{v}}_{\omega,ij} \right] \hat{\mathbf{v}}_{\omega,ij}}_{\text{translational slip}}, \label{relative_velocity}
\end{equation}
where $\hat{\mathbf{v}}_{\omega,ij} = \mathbf{v}_{\omega,ij}/|\mathbf{v}_{\omega,ij}|$ is the normalized rotational slip velocity, with $\mathbf{v}_{\omega,ij}=\frac{\sigma}{2} (\boldsymbol{\omega}_i + \boldsymbol{\omega}_j) \times \hat{\mathbf{r}}_{ij}$. The tangential friction force $\mathbf{f}_{t,ij}$ facilitates energy transfer between rotational and translational degrees of freedom. Note that both $\mathbf{f}_{t,ij}$ and $\mathbf{f}_{d,ij}$ act as the source of the energy dissipation which induces phase separation as shown later.

Our numerical simulations are performed in a periodic box of size $L$ (spinner number density $\rho_0 = N/L^2$) with time step $\Delta t = 0.005 \tau_0$, where $\tau_0 = m/\gamma_t$ sets the timescale and $\epsilon=\sigma^2\gamma_t^2/m$ sets the energy scale. Other model parameters are fixed as {$I = m\sigma^2/8$}, $\gamma_r = 0.3 m \sigma^2/\tau_0$, $k_e = 50 m/(\tau_0^{2} \sigma^{1/2})$, $k_d = 3\tau_0 k_e$ and $k_t = 10 m/\tau_0$.

\begin{figure}[!bhtp]
	\resizebox{70mm}{!}{\includegraphics[trim=0.0in 0.0in 0.0in 0.0in]{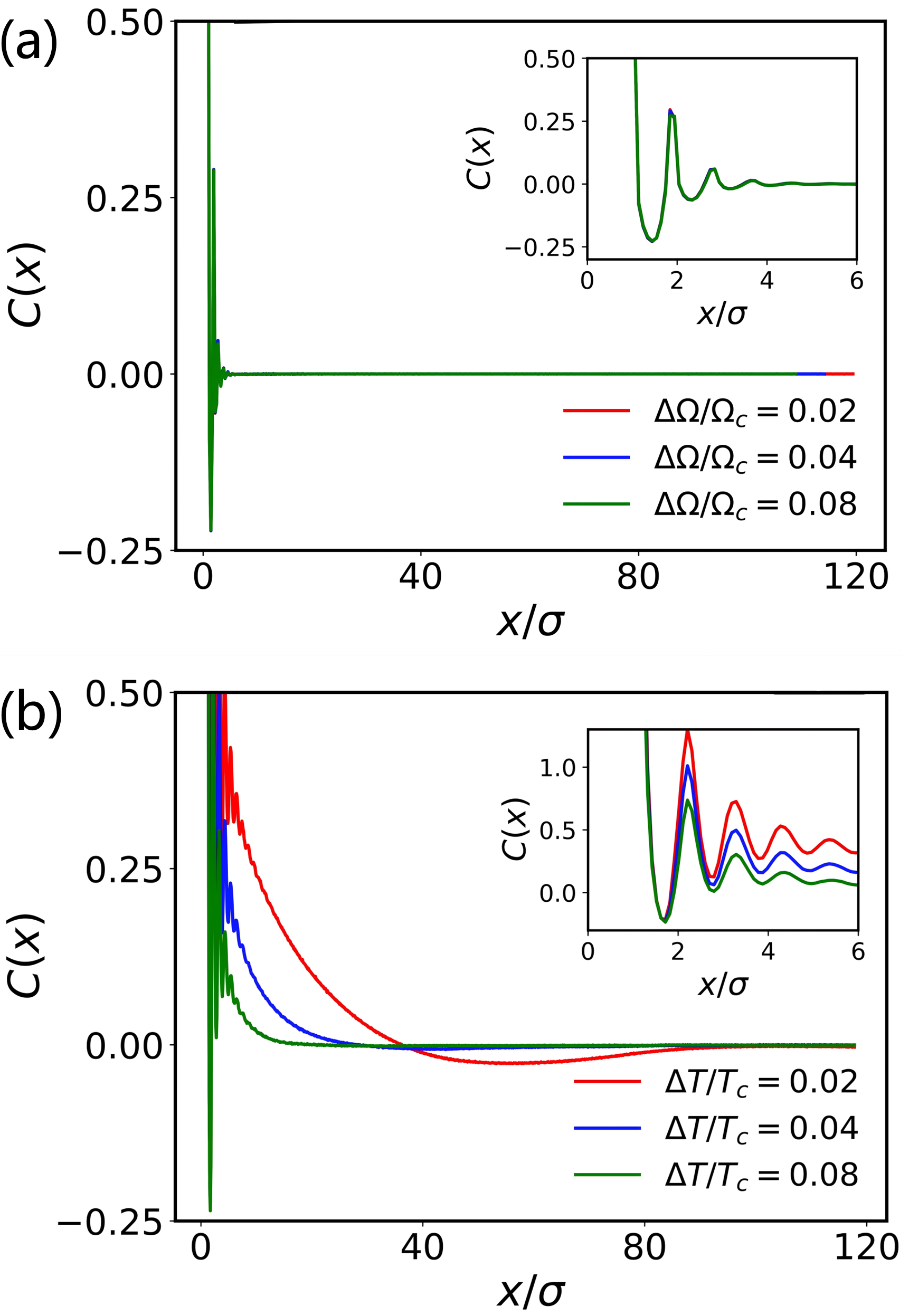} }
\caption{{Comparison of the pair correlation functions between 2D HU fluids (a) and 2D equilibrium Lennard-Jones fluids (b) near the LG critical point, where $\Omega_c$ ({$T_c$}) is the critical torque (temperature) and $\Delta \Omega$ ($\Delta T$) is the distance from it.}    }\label{MainFigCor}
\end{figure}

\section{Dissipation-Induced Phase Separation}
The system is a typical driven-dissipative system: the energy injected through rotational torque $\boldsymbol{\Omega}$ is dissipated by collisions and friction in both translational and rotational degrees of freedom. Under low density or weak driving ($\boldsymbol{\Omega}\to0$) conditions, dissipation dominates. The system thus falls into absorbing states where each spinner rotates in its fixed place (Fig.~\ref{MainFigSpinner}(d)). In contrast, strong driving or high density sustains a homogeneous active phase marked by persistent collisions (Fig.~\ref{MainFigSpinner}(e)) and a HU fluid state with structure factor scaling $S(q)\sim q^2$ as $q\to 0$~\cite{Lei2019h} {(Fig.~\ref{MainFigSpinner}(c))}. Earlier studies showed that for systems with non-dissipative collisions, the absorbing phase transition belongs to the universality class of conserved directed percolation (CDP)~\cite{Lei2019h}.

Our work extends the above framework by introducing dissipation during collisions, which has a pronounced effect on the density field: high-density regions exhibit amplified dissipation due to frequent interparticle collisions, decreasing local kinetic energy and pressure compared to low-density regions. This imbalance would generate a negative compressibility that destabilizes the density field~\cite{clewett2012emergent,clewett2016minimization,herminghaus2017,fullmer2017clustering, mandal2019motility}, ultimately triggering LG phase separation at sufficiently large $\Omega$, resulting in the coexistence of a gas phase (high kinetic temperature) with a liquid phase (low kinetic temperature) as illustrated in Fig.~\ref{MainFigSpinner}(a, f, g).  We term this phenomenon dissipation-induced phase separation (DIPS), which is different from the phase separation due to local jamming~\cite{liu2023local}, odd response~\cite{nguyen2014emergent,ding2024odd}, hydrodynamic interactions~\cite{yeo2015collective,shen2023collective}. 
A similar mechanism has also been reported in non-equilibrium granular gas systems~\cite{clewett2012emergent}. The resultant phase diagram in ($\rho$, $\Omega$) space (Fig.~\ref{MainFigSpinner}(b)) delineates three distinct regimes: absorbing state (cyan), HU fluids (red), and LG coexistence (green).

In Fig.~\ref{MainFigSpinner}(c), we show $S(q)$ for systems at different $\Omega$ along a trace crossing the LG critical point ({$\rho_c= 0.70\sigma^{-2},~\Omega_c=3.4\epsilon$}). Below the critical point, the active phase maintains hyperuniformity with $S(q \to 0) \sim q^2$~\cite{PhysRevLett.118.020601}. Above the critical point, phase separation manifests {through $S(q) \sim q^{-3}$}, aligning with Porod's law~\cite{kostorz2001phase,CriDyn}. Remarkably, at the critical point, we observe non-conventional finite fluctuations ($S(q \to 0) \sim \text{const.}$), contrasting the divergent   fluctuations $S(q \to 0) \sim q^{-2}$ of conventional LG criticality~\cite{CriDyn}. The spatial pair correlation functions also exhibit a short-range correlation near criticality, distinct from a quasi-long-range correlation in equilibrium fluids (see Fig.~\ref{MainFigCor}).  
Simulation details for equilibrium fluids can be found in SM~\cite{supmat} Sec.~II.
These non-conventional fluctuations and correlations suggest that hyperuniformity fundamentally reshapes LG criticality.
Furthermore, in Fig.~\ref{MainFigGFDR}(a) we measure the kinetic energy spectrum 
$T_t( q) = \langle \vert v(q) \vert^2 \rangle/2$
of the active spinner system (see \add{simulation} details in SM~\cite{supmat} Sec. III). We confirm that the kinetic temperature obeys the power law $T_t(q)\sim q^2$, in agreement with the general FDR Eqs.~(\ref{GFDR}-\ref{GFDR2}).

\begin{figure*}[thbp]
	\resizebox{180mm}{!}{\includegraphics[trim=0.0in 0.0in 0.0in 0.0in]{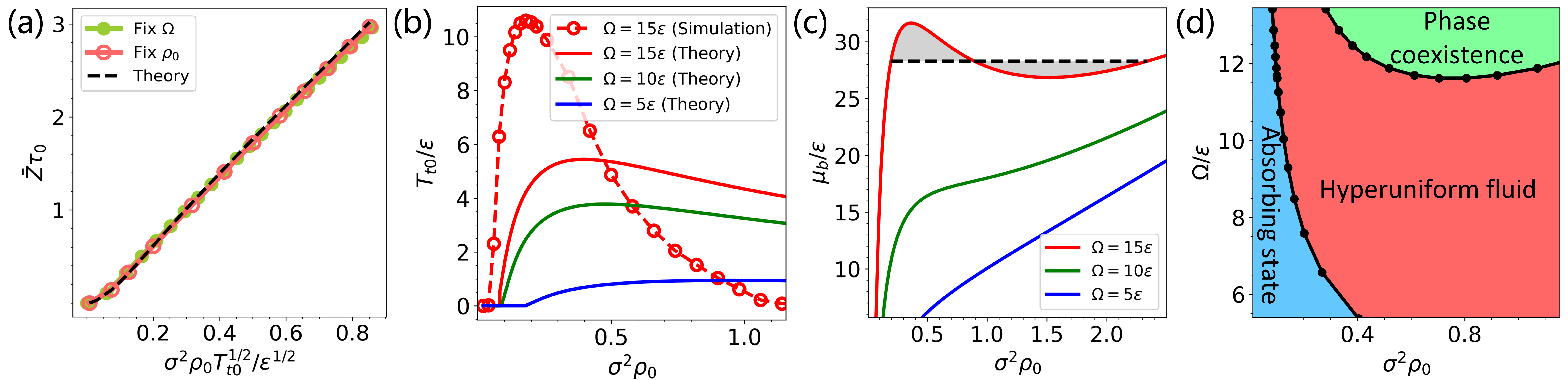} }
\caption{Hydrodynamic theory of active spinner: (a) Theoretical prediction of average collision frequency$\Bar{Z}$ as a function of  $T_{t0}$ or $\rho_0$ compared with simulation results under fixed  $\rho_0$  (red line)  or rotational torque $\Omega$ (green line). (b) Translational kinetic energy as a function of $\rho_0$.  (c) Effective bulk chemical potential $\mu_b$ as a function of $\rho_0$, where a Maxwell construction can be done for the $\Omega=15 \epsilon$. (d) Theoretical phase diagram similar to Fig. 1(b). }\label{MainFigHydro}
\end{figure*}

\section{Hydrodynamic theory of active spinner fluids}
To characterize the phase behavior of active spinner systems, we develop a hydrodynamic field theory describing the evolution of the density field  $\rho(\mathbf{r},t)$, velocity field $\mathbf{u}(\mathbf{r},t)$ and the rotational/translational kinetic energy (temperature) fields $T_{r/t}(\mathbf{r},t)$ of active spinners (see Appendix B for detailed derivations)
\begin{align}
    \frac{\partial \rho}{\partial t} &= - \nabla\cdot (\rho \mathbf{u}),\label{contiEq}  \\
   {m \rho}  \frac{ \mathcal{D} \mathbf{u}}{\mathcal{D} t} &=  - \nabla P +  {{\eta}^*} \nabla^2 \mathbf{u} +  \zeta_0 \nabla(\nabla\cdot \mathbf{u}) \nonumber  -  \gamma_t  \rho\mathbf{u}  + \nabla \cdot \boldsymbol{\sigma}_{ \mathbf{u}},  \nonumber \\ \label{NSEq} \\
    \frac{\mathcal{D}T_t}{\mathcal{D}t} &= D_t\nabla^2 T_t + s_1 \Bar{Z} T_r - s_2\Bar{Z} T_t - \frac{2\gamma_t }{m} T_t + \eta_t(\mathbf{x},t),  ~~~~~~~~ \label{kineticEqn1}  \\
\frac{ \mathcal{D} T_r}{\mathcal{D} t}  &= D_r\nabla^2 T_r   + s_3 \Bar{Z} T_t  - s_4\Bar{Z} T_r - \frac{2\gamma_r }{I} T_r  +  \frac{2\gamma_r T_{ss}^{1/2} }{I} T_r^{1/2}  \nonumber \\ 
&  + \eta_r(\mathbf{x},t), \label{kineticEqn2}
\end{align}
where $ \frac{ \mathcal{D} }{\mathcal{D} t} =  \left( \frac{\partial }{\partial t} + \mathbf{u}\cdot\nabla \right) $ is the material derivative.  $P$ is the pressure field. {$\eta^*=\eta_0 + \eta^o \boldsymbol{\epsilon}\cdot$}, where  $\eta_0$ is the normal shear viscosity, $\eta^o$ is the odd viscosity for active spinner systems with $\boldsymbol{\epsilon}$  the $\pi/2$ clockwise rotation matrix~\cite{Han2021,lou2022odd,fruchart2023odd,ding2024odd}. $\zeta_0$ is the bulk viscosity. $D_r$ and $D_t$ in  \add{Eqs.~(\ref{kineticEqn1}-\ref{kineticEqn2})} are the diffusion constants for the two kinetic energy fields.  \add{${\gamma_t }T_t$} and  ${\gamma_r }T_r$  terms  represents \add{the translational} and rotational friction dissipation, respectively.  In Eq.~(\ref{kineticEqn2}), the $ T_{ss}^{1/2}T_r^{1/2}$ term represents the driving power of the constant torque with $T_{ss} = I\Omega^2/(2\gamma_r^2)$.  The $\Bar{Z} T_r$ and $\Bar{Z}T_t$ terms in  \add{Eqs.~(\ref{kineticEqn1}-\ref{kineticEqn2})} account for  energy gain/loss  due to collisions in the translational and rotational \add{degree of freedom}, where the average collision frequency {$\Bar{Z} =  (1-e^{- l_d/\lambda})\bar{v}/{\lambda}$} with the mean free path $\lambda = 1/(\sqrt{2 \pi}\rho\sigma )$~\cite{cutchis1977enskog,visco2008collisional}.  
The prefactor $(1-e^{-l_d/\lambda})$ is the next-collision survival probability accounting for the damping of the translational velocity (see Appendix B). $s_1$, $s_2$, $s_3$  $s_4$ are \add{coefficients describing} the exchange of kinetic energy  between the rotational and translational degrees of freedom~\cite{luding1998homogeneous}, which can be measured in simulations (SM~\cite{supmat} Sec. I). 
Note that the $s_4$ term also accounts for the energy dissipation effect arising from inelastic collisions. $\eta_r(\mathbf{x},t)$, $\eta_t(\mathbf{x},t)$ and $\boldsymbol{\sigma}_{\mathbf{u}}(\mathbf{x},t)$ are the noise terms. Since the system enters the absorbing state when $T_t=0$, the noise $\eta_t(\mathbf{x},t)$ satisfies $\langle \eta_t(\mathbf{x},t) \eta_t(\mathbf{x}',t') \rangle=F_t(T_t)\delta^d(\mathbf{x}-\mathbf{x}')\delta(t-t')$ with $F_t(0)=0$. The same requirement applies to $\eta_r(\mathbf{x},t)$ and $\boldsymbol{\sigma}_{\mathbf{u}}(\mathbf{x},t)$.

In the long-wavelength limit, the density field $\rho$ emerges as the sole slow mode. This permits the neglect of material derivative terms in the hydrodynamic equations (see Appendix C). In the homogeneous state, $\rho(\mathbf{r},t)=\rho_0$, $T_t(\mathbf{r},t)=T_{t0}$, $T_r(\mathbf{r},t)=T_{r0}$,$P(\mathbf{r},t)=P_0$, and $\mathbf{u}(\mathbf{r},t)=0$, the kinetic energies at the mean-field level satisfy
\begin{eqnarray}
    0 &=&   s_1 \Bar{Z} T_{r0} - s_2\Bar{Z} T_{t0} - \frac{2\gamma_t }{m} T_{t0} . \label{EffkineticEqn3}\\
     0 &=&   s_3\Bar{Z} T_{t0} - s_4 \Bar{Z} T_{r0}  +  \frac{2\gamma_r }{I} (T_{ss}^{1/2}T_{r0}^{1/2}-T_{r0})       \label{EffkineticEqn2}   
\end{eqnarray}
The numerical solution of Eqs.~(\ref{EffkineticEqn3}-\ref{EffkineticEqn2}) gives $T_t(\rho_0)$. Fig.~\ref{MainFigHydro}(b) shows $T_t(\rho_0)$ under three characteristic driving torques $\Omega$, revealing a non-monotonic density dependence at large driving torque $\Omega$, in agreement with molecular dynamics simulations (red dashed symbols). The critical point of the absorbing state ($T_{t} = 0$) can be unambiguously identified in Fig.~\ref{MainFigHydro}(b). The theoretically predicted phase boundary of the absorbing state transition shown in Fig.~\ref{MainFigHydro}(d) is qualitatively consistent with active spinner systems.
Analysis within this hydrodynamic framework (see Appendix D) establishes that the absorbing transition belongs to the CDP universality class~\cite{henkel2008non}.

By expanding the pressure $P$ in terms of the local fields and their derivatives, we can derive the relation between pressure $P$, $\rho$ and $T_t$,
\begin{eqnarray}\label{perturbation_2} 
P (\rho,\nabla^2\rho)  =   \frac{1}{\Bar\chi_T}  \rho + \beta_V  T_t(\rho) - K_{\rho} \nabla^2  \rho+\cdots
\end{eqnarray} 
The compression coefficient $\Bar{\chi}_T$ and pressure coefficient $\beta_V$ can be estimated based on the equation of state for hard disks derived from scaled-particle theory~\cite{helfand1961theory}  $\pi \sigma^2 P=4T_{t0} y / (1-y)^2$ with $y=\pi\rho_0\sigma^2/4$.
Near the critical point, this gives $1/\Bar{\chi}_T = T_{t,c}(1+y^*)/[(1-y^*)^3]$ and $\beta_V=4y^*/[(\pi \sigma^2)(1-y^*)^2]$\add{, where $T_{t,c}= 0.4\epsilon$}. The term $K_{\rho} \nabla^2 \delta \rho$  with $K_{\rho}>0$ provides the positive surface tension in the formation of the interface.  
{Substituting relations $T_t(\rho)$ and $P(\rho,\nabla^2\rho,T_t)$ into Eqs.~(\ref{contiEq}-\ref{NSEq})}, we can derive a Model B-like equation (see details in Appendix C)
\begin{eqnarray}  \label{RCH-AA}
    \frac{\partial \rho }{\partial t} = {\gamma_t}^{-1}\sigma^{-2}\nabla^2\left( \mu_b - \kappa_{\rho} \nabla^2 \rho  \right)    - {\gamma_t}^{-1} \nabla^2 \sigma_{\parallel, u} ~~
\end{eqnarray}
where {$\kappa_{\rho}=\sigma^2 K_{\rho}$ and} $\sigma_{\parallel, u}$ denotes the longitudinal component of random noise  $\boldsymbol{\sigma}_{ \mathbf{u}}$, {and noise variance $\langle\sigma_{\parallel, u}(\mathbf{x},t)\sigma_{\parallel, u}(\mathbf{x}',t')\rangle=2\rho_0 \nu_{\parallel} T_t \delta^d(\mathbf{x}-\mathbf{x}')\delta(t-t')$, where $\nu_{\parallel}$ is the longitudinal kinematic viscosity~\cite{gross2010thermal,Lei2019h}.  Note that higher order active flux terms that break detailed balance at the field level, such as $\nabla^2\left( \nabla \rho \right)^2,~\nabla\cdot \left[ (\nabla^2\rho)\nabla\rho \right]$, generically exist in active matter systems~\cite{wittkowski2014scalar,cates2025active}. However, in our system, they can be safely neglected in the long-wavelength limit due to the suppressed density fluctuations in HU fluids (see Appendix E Sec. 3). Without the influence of these active flux terms, the deterministic terms $\mu_b(\rho)$ in the field equation are identical to those of an equilibrium system. Thus $\mu_b(\rho)$ can be formally written as effective bulk chemical potential}
\begin{eqnarray}
     &&\mu_b(\rho)  
     = \frac{\partial f_b}{\partial \rho}
     = \frac{\sigma^2}{ \Bar\chi_T}  \rho+  \sigma^2\beta_V T_t(\rho) \label{chem_potent}
\end{eqnarray}
The corresponding effective free energy functional is 
\begin{eqnarray}\label{FreeEnergy}
    \mathcal{F}[\rho] = \int d^dx~ \left[f_b(\rho) + \frac{1 }{2}\kappa_{\rho}|\nabla \rho|^2\right]
\end{eqnarray}
A similar effective free energy has also been introduced to {phenomenologically} describe motility-induced phase separation (MIPS) in active Brownian particle systems~\cite{speck2014effective,takatori2015towards}. It should be emphasized that while odd viscosity emerges intrinsically in active spinner fluids, the hydrodynamic theory reveals that it has no effect on the LG critical phenomena \add{(Appendix C)}.

\begin{table*}[!htb] 
{
\begin{tabular}{c c c c c c c c c }
\hline \\[-2.0ex]
  & ~~~$S(q)$~~~ &~~~$U_4^*$~~~ & ~~~$C(x)$~~~ & ~~~$\chi_{\rho}(\tau,L)$~~~ &~~~$c_v(\tau,L)$~~~&  ~~~$t_w$~~~ &\\[1.5ex]
\hline 
\\[-1.5ex]

{MF Ising ($d > 4$)}~~  & $q^{-2}$ & \add{$0.457$}  & {$x^{2-d}$} & $L^{2}S_{\rho}(\tau L^2)$ & $C_v(\tau L^2)$ 
& $\tau^0$ \\ [1.5ex]

{4D Ising model}~~  & $q^{-2}$ & \add{{$0.45$}}  &  $ {x^{-2}}$ & $L^{2}\ln^{\frac{1}{2}}LS_{\rho}(\tau L^2\ln^{\frac{1}{6}}L)$ & $\ln^{\frac{1}{3}}LC_v(\tau L^2\ln^{\frac{1}{6}}L)$ & $\tau^0$
    \\ [1.5ex]

{2D Ising model}~~  & \add{$q^{-7/4}$} & \add{$0.856$}  &  $ x^{-1/4}$ & $L^{7/4}S_{\rho}(\tau L)$ & $\ln L~ C_v(\tau L)$ & $\tau^0$
    \\ [1.5ex]

{2D HU fluids}~~ 
    & $const.$
    & ${1}/{3}$
    & $\delta^2(\mathbf{x})-\tau [\frac{3}{2}\delta\overline{u}_0\ln \Lambda x]^{-\frac{1}{3}} $ 
    & $\ln^{\frac{1}{2}}LS_{\rho}(\tau L^2\ln^{\frac{1}{6}}L)$ 
    & $f(\tau) + c_0 L^{-\lambda} + c_1 \tau L^{2 - \lambda}$  
    & $\tau^{-2}$
        \\ [1.5ex]

\hline \\
\end{tabular}
\caption{ Comparison of the critical behaviors of 2D HU fluids and the Ising model. Data for 2D, 4D and the mean-field Ising model are taken from Refs.~\cite{onsager1944crystal,luijten1997interaction,aktekin2001finite,kenna2006self,
lv2021finite,li2024logarithmic,selke2006critical}.  $t_w$ for the noiseless MF Ising model satisfies a different scaling law Eq.~(\ref{waitingTime}). } \label{Tab_2}
}
\end{table*}

% ~\cite{kamieniarz1993universal}

Fig.~\ref{MainFigHydro}(c) displays $\mu_b(\rho)$ under varying $\Omega$ based on Eq.~(\ref{chem_potent}), revealing non-monotonicity that signals LG instability. This behavior originates directly from the non-monotonicity of $T_t(\rho)$ shown in Fig.~\ref{MainFigHydro}(b). 
Given that the deterministic terms in Eqs.~(\ref{RCH-AA}-\ref{FreeEnergy}) are identical to those in equilibrium systems, we employ the Maxwell construction to obtain an estimate of the coexisting gas and liquid densities~\cite{speck2014effective,takatori2015towards}. These results are plotted in Fig.~\ref{MainFigHydro}(d). We note that the Maxwell construction is generally inaccurate for non-equilibrium phase separation~\cite{wittkowski2014scalar}. 
Nevertheless, the qualitative agreement between the theoretical phase diagram Fig.~\ref{MainFigHydro}(d) and the simulation phase diagram Fig.~\ref{MainFigSpinner}(b) suggests that our theory and approximation successfully capture the essential physics of active spinner systems.

%Since the magnitude of the critical external torque is related to the microscopic details of the system, the long wave approximation theory can only qualitatively characterize the macroscopic behavior of the system without accurately predicting the critical external torque.
 
\section{Renormalization-Group Analysis}

Although the deterministic terms in Eq.~(\ref{RCH-AA}) are equilibrium-like, the noise term, stemming from reciprocal collision interactions, satisfies 
center-of-mass conservation and is therefore intrinsically non-equilibrium. Without loss of generality, we adopt the Ginzburg-Landau-like form~\cite{hohenberg1977theory} of Eq.~(\ref{RCH-AA}), i.e.,
\begin{eqnarray}   \label{CMC-modelB-s}
    \frac{\partial \psi}{\partial t}
    =  \nabla^2  \left({ r\psi + \frac{u}{6} \psi^3  - \kappa\nabla^2 \psi  }\right)
   +  {\zeta}(\mathbf{x},t)
\end{eqnarray} 
with noise correlations $\langle{ \zeta(\mathbf{x},t)\zeta(\mathbf{x}',t') }\rangle = 2D \nabla^{4}\delta^d(\mathbf{x}-\mathbf{x}')\delta(t-t')$, where $D$ is the noise strength. The scaling law of the structure factor $S(q) $ can be obtained as (see Appendix E Sec. 2 and Sec. 3)
{\begin{eqnarray}   \label{S_k}
S(q)  = \langle\psi_q \psi_{-q}\rangle  \sim
\left\{ \begin{aligned}
& q^2 ,~~ & r > r_c \\
& q^{\eta}~~ & r=r_c \\
\end{aligned} 
 \right. 
\end{eqnarray} }
where $r=r_c$ corresponds to the critical point of Eq.~(\ref{CMC-modelB-s}) {and $\eta$ is the anomalous dimension, which we will discuss later.}
This prediction agrees with simulation results of HU spinner fluids.

Under the scaling transformation  $\mathbf{x} \to  b\mathbf{x'} $, the time, field and noise are changed as  $t \to b^z t'$, $\psi \to b ^{\chi} \psi'$ and $\zeta\to b^a \zeta'$, where $z$ and $\chi$ are the scaling dimension of $t$ and $\psi$.   Eq.~(\ref{CMC-modelB-s}) becomes
\begin{eqnarray}
    \frac{\partial \psi'}{\partial t'}
    &=&
    \nabla'^2\left({
        b^{z-2} r\psi' - b^{z-4} \kappa\nabla'^2 \psi' + b^{2\chi+z-2} \frac{u}{6} \psi'^3
    }\right)  \nonumber \\
   &&  + b^{a-\chi+z}  \zeta'( b \mathbf{x}',b^z t' ) .
\end{eqnarray}
And 
\begin{align}
    \langle{ \zeta'( b\mathbf{x}_1', b^zt_1')\zeta'( b\mathbf{x}_2', b^zt_2') }\rangle
    &= 
    2b^{-2a-4-d-z}D\nonumber\\
    &\times \nabla'^4\delta^d(\mathbf{x}_1'-\mathbf{x}_2')\delta(t_1'-t_2').
\end{align}
In the vicinity of the Gaussian point ($r=0$), the invariance of Eq.~(\ref{CMC-modelB-s}) requires
\begin{eqnarray}
b^{z-4}=1,~~b^{a-\chi+z}=1,~~b^{-2a-4-d-z}= 1
\end{eqnarray}
which leads to $z=4$ and  $\chi=-d/2$. Thus when $d>2$,  the nonlinear term associated with $u$ becomes irrelevant under the renormalization-group (RG) transformation $u \to u'=b^{2-d}u$, making the upper critical dimension $d_c=2$ different from 4 for the Ising universality class. It should be noted that a recent work discussed the effect of spatially or temporally correlated noise on the upper and lower critical dimension of general $O(n)$ model and related phase transitions~\cite{ikeda2023correlated}. It was suggested that the upper critical dimension is $d_c=3$ for a system with $n=1$ (the case studied here), which is different from our prediction. In the next section, we will give clear evidence that the upper critical dimension is indeed $d_c=2$.

Next we derive the flow equations for couplings $r$ and $u$ using Wilson's momentum shell RG approach {below the upper critical dimension} (see Appendix E Sec. 4)~\cite{CriDyn}. We obtain the critical exponents to the lowest non-trivial order, which are formally the same as those of the Ising universality class despite the significant change of $d_c$, namely,
\begin{equation}
\nu \approx \frac{1}{2}+\frac{\varepsilon}{12},~ \beta \approx \frac{1}{2}-\frac{\varepsilon}{6},~\delta\approx3+\varepsilon,~\gamma \approx 1+\frac{\varepsilon}{6},~ {\eta\approx \frac{\varepsilon^2}{54}},~z\approx4
\end{equation}
where $\varepsilon = d_c - d=2-d$.
{Nevertheless, the critical behaviors of $d$-dimension HU fluids are still in stark contrast to the $(d+2)$-dimension Ising model. These differences are summarized in Table~\ref{Tab_2} and will be explained in the rest of the paper.}
{For example, dimensional analysis of Eq.~(\ref{PrimeScaling}) yields $S(q)\sim q^{\eta}$ while for the Ising model $S(q)\sim q^{-2+\eta}$. 
This implies that the $d<d_c$ system remains hyperuniform even at the critical point.}

\begin{figure*}[!bhtp]
	\resizebox{175mm}{!}{\includegraphics[trim=0.0in 0.0in 0.0in 0.0in]{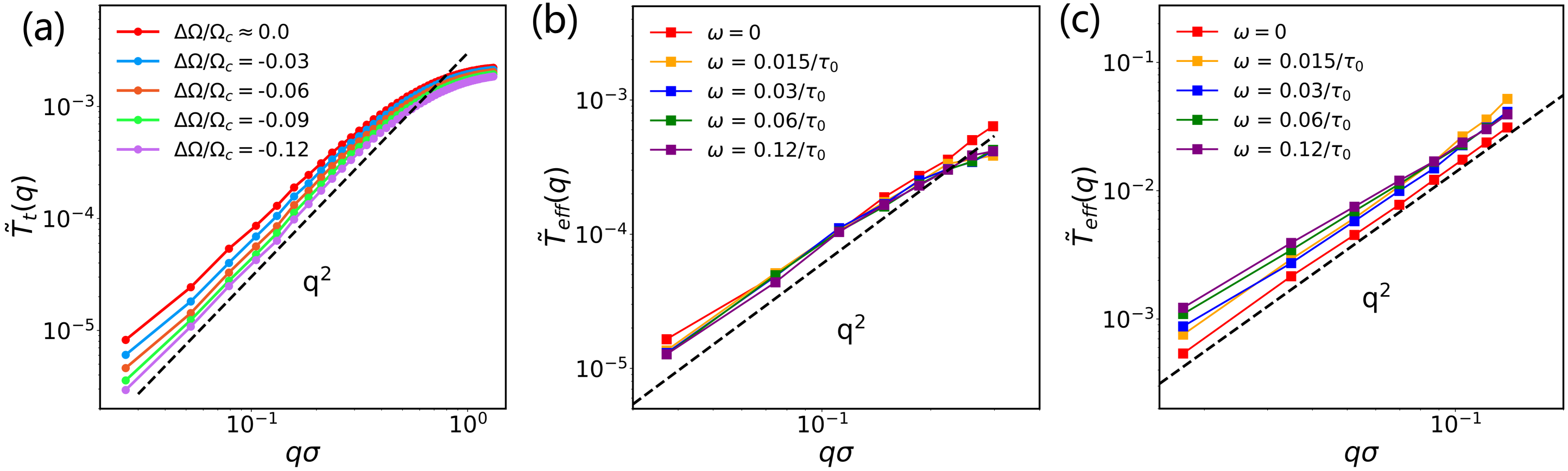} }
\caption{  { (a) Kinetic energy spectrum $T_t( q) = \langle \vert v(q) \vert^2 \rangle/2$ of active spinner systems. (b,c) Effective temperature of spinner systems at {$\Delta \Omega/\Omega_c=-0.12$} (b) and 2D stochastic field at {$\Delta r/r_c=-0.16$} (c). $k_B T_{\text{eff}}(q)$ is defined by the generalized FDR Eq.~(\ref{GFDR}) and Eq.~(\ref{GFDRstatic}).}
}\label{MainFigGFDR}
\end{figure*}

{Moreover, based on our dynamic field theory (see Appendix E Sec.~2), we obtain the short-range asymptotic behavior of the pair correlation function of HU fluids near LG critical point instead of quasi-long range as expected for the Ising universality class~\cite{kenna2006self,ElementsCri}, i.e.,}
\begin{eqnarray}\label{MeanFieldCorrelation}
     C(x)  ~ \sim  ~ \frac{\delta^d(\mathbf{x})}{x^{\eta}} - \frac{D\tau^{2\nu}}{x^{d-2+\eta}},\ \ \  x\ll \xi.  \label{correlation_func}
\end{eqnarray} 
{where $\tau =(r -r_c)$ and $\xi=\tau^{-\nu}$ is the correlation length.
Note that the first $\delta(x)$ function term leads to a peak near $x=0$, while the second term gives a negative contribution whose magnitude is controlled by the distance to the critical point, which vanishes at the critical point. 
This apparent short-range correlation indicates the density field is structureless and homogeneous at LG criticality, consistent with the behaviors of $S(q)$.
Note that when $d = d_c$, the second term requires a logarithmic correction (see Appendix E Sec. 5).}

We further derive the response (susceptibility) function $\chi(x)$, i.e., the response of the field at distance $x$ to a perturbation applied at the origin
\begin{eqnarray}\label{Sus}
    \chi(x) 
    \sim 
    \dfrac{1}{x^{d-2+\eta'}},\ \ \ 
    x\ll \xi
\end{eqnarray}
where $\eta'$ is the anomalous dimension associated with the response function~\cite{Young2020}. Near the critical point $\tau=0$, the correlation length $\xi$ diverges and the response function exhibits a power-law decay similar to equilibrium fluids. This suggests that HU fluids, {despite being  homogeneous and calm}, are highly susceptible at the critical point.

In the Ising model, magnetization fluctuations ${\langle M^2\rangle - \langle M\rangle^2}$ connect to thermodynamical susceptibility $\chi_m$ through the fluctuation-dissipation theorem
$
\chi_m = \frac{\langle M^2\rangle - \langle M\rangle^2}{Nk_B T} \sim \tau^{-\gamma}
$
, where $M$ represents magnetization. Analogously, for equilibrium fluids near criticality, the density fluctuations in sub-boxes of liquid/gas phase $\chi_\rho = \langle(N_s - \langle N_s\rangle)^2\rangle / \langle N_s\rangle$, are related to the compressibility $\chi_c$ via 
$
\chi_c ={\chi_\rho}/{k_B T} \sim \tau^{-\gamma'}
$, which satisfies conventional finite-size scaling~\cite{luijten1997interaction,aktekin2001finite, li2024logarithmic}:
\begin{equation} \label{compressibility_finite_size}
\chi_c(\tau, L) = \left.\frac{\partial  \Delta\psi_{lg}  }{\partial h}\right|_{h\to0} = L^{\gamma'/\nu} {S}_c(\tau L^{1/\nu})
\end{equation}
with the scaling function  ${S}_{c}(x) \sim x^{-\gamma'}$ for $x \gg 0$ and ${S}_{c}(x) \sim  {const.}$ as $x \to 0$. For HU fluids, dimensional analysis reveals that the density fluctuation $\chi_\rho$ satisfies a {non-conventional} finite-size scaling (see Appendix E Sec. 3)
\begin{eqnarray} \label{S0_finite_size}
\chi_{\rho}(\tau,L) 
    =
    L^{\gamma/\nu-2}{S}_{\rho}(\tau L^{1/\nu})
\end{eqnarray}
the scaling function  ${S}_{\rho}(x)$ is similar to ${S}_{c}(x)$. Consequently, above the critical temperature ($\tau \gg 0$), the system exhibits suppressed density fluctuations scaling $ \chi_{\rho} (L) \sim L^{-2}$ which is characteristic of HU fluids, while near criticality ($\tau \to 0$), $\chi_\rho(L) \sim L^{\gamma/\nu-2} \sim {const.}$ emerges  as a result of $\gamma/\nu=2$, signalling {finite} fluctuations. 
It should be noted that this finite-size density fluctuations scaling is different from the scaling of density fluctuations usually used to define hyperuniformity~\cite{Torquato2018}, since the observation windows used here should increase with system size. Eqs.~(\ref{compressibility_finite_size}-\ref{S0_finite_size}) again suggest that HU fluids are  calm yet highly susceptible, fundamentally different from the Ising universality class.

In many non-equilibrium  systems, the large-scale behaviors can be effectively described by an equilibrium  theory~\cite{speck2014effective, takatori2015towards,Han2017,Gnan2022}. 
Such systems satisfy the effective FDR  $\chi(x) =C(x) /(k_B T_{\rm eff})$ or $\chi_c = \chi_\rho/(k_B T_{\rm eff})$, from which a single effective temperature $T_{\rm eff}$ can be defined~\cite{CriDyn,palacci2010sedimentation,
petrelli2020effective,Han2017,hecht2024define}. 
The complete difference between $\chi(x)$ and $C(x)$, along with the disparity between $\chi_c$ and $\chi_\rho$, indicates a fundamental violation of the conventional FDR for HU fluids at the critical point.  
However, HU fluids still obey the generalized FDR i.e., Eq.~(\ref{GFDR}) with a $q$-dependent effective temperature $k_BT_{\rm eff}= Dq^2$.  From  Eq.~(\ref{GFDR}), the static generalized FDR ($\omega=0$) can also be obtained (see Appendix A)
\begin{align}
    \chi(q) =\frac{S(q)}{k_BT_{\text{eff}}(q)}\label{GFDRstatic}
\end{align} 
It should be emphasized that Eq.~(\ref{GFDR}) is a generic result for non-equilibrium systems with center-of-mass conservation, independent of the interaction type and whether the system is at the critical point. This special property is the fundamental origin of non-conventional critical behaviors of HU fluids, whose calm yet highly susceptible nature can be understood as $T_{\rm eff} \rightarrow 0$ at long wavelengths. Note that at the critical point $(r = 0, u = 0)$, Eq.~(\ref{CMC-modelB-s}) appears to fall back into a center-of-mass  conserved equilibrium  diffusive system~\cite{han2024scaling}
\begin{equation}\label{nabla4}
    \frac{\partial \psi}{\partial t}
    =   - \nabla^4   \psi  
   +  \zeta(\mathbf{x},t)
\end{equation}
However, the response (or dissipation) of the system is still determined by Eq.~(\ref{CMC-modelB-s}) making the LG critical point impossible to map to an equilibrium system, despite the similar Gaussian fluctuations in the two systems.

\begin{figure*}[bhtp]
	\resizebox{180mm}{!}{\includegraphics[trim=0.0in 0.0in 0.0in 0.0in]{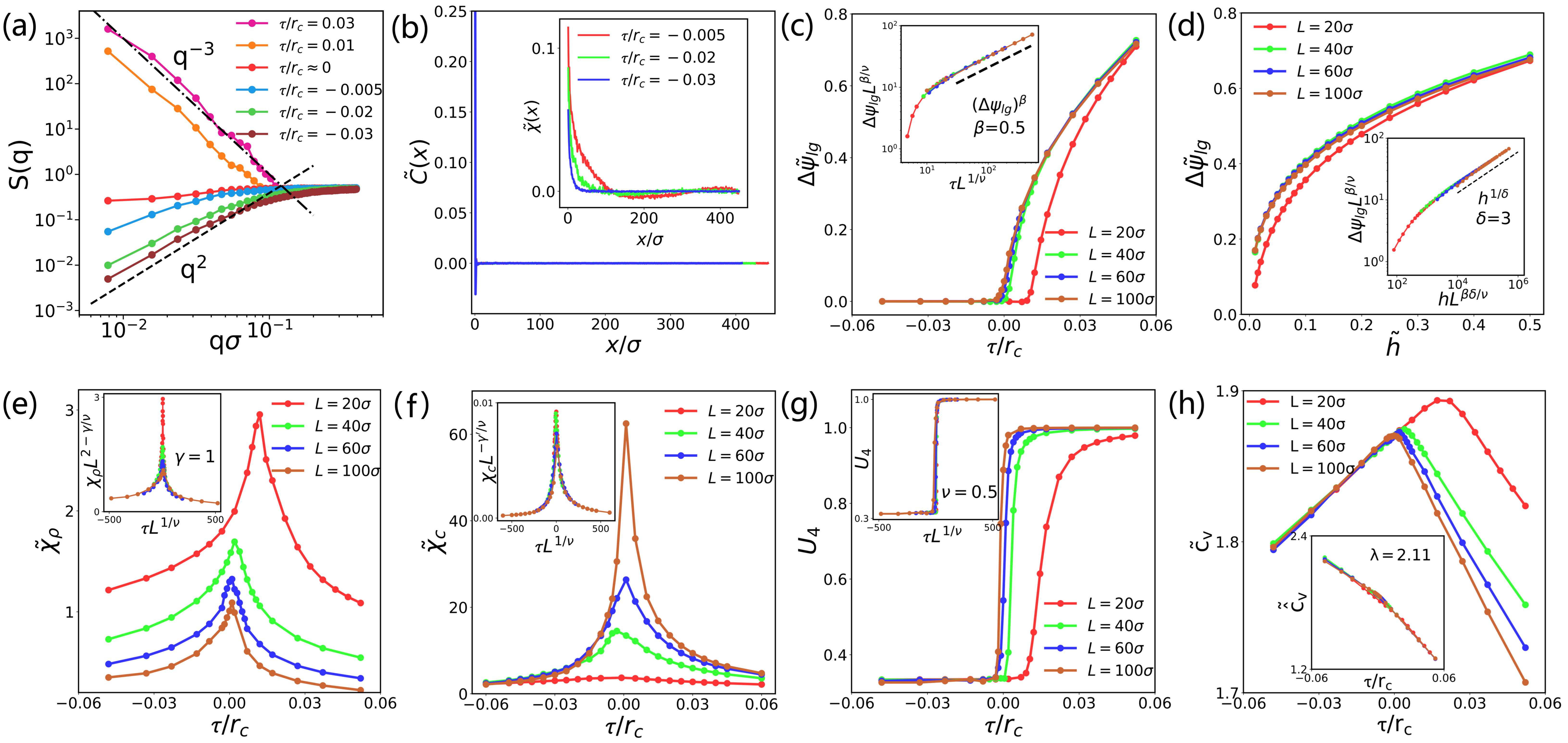} }
\caption{ Simulation results of 2D stochastic field. (a) structure factor for system at different distances from the critical point $\tau$. (b) correlation function and response function near the critical point \add{(Fig.~S5 provides a log-linear plot version in SM~\cite{supmat})}. Finite-size scaling analysis for (c) order parameter, (d) response to an external field, (e) density fluctuations, (f) compressibility, (g) Binder cumulant and (h) energy fluctuation near the LG critical point, where $\hat{c}_v(\tau,L) = {c}_v(\tau,L) - c_0 L^{-\lambda} - c_1 \tau L^{1/\nu - \lambda}$. }\label{MainFigFSS}
\end{figure*}

\section{2D stochastic field simulation} 
To validate our theoretical predictions, we perform large-scale stochastic field simulations of Eq.~(\ref{CMC-modelB-s}). 
In actual calculations, fluctuations lead to a downward shift of the critical temperature $r_c$~\cite{CriDyn}. 
Through finite-size scaling analysis, we determine $r_c=-11.97\pm0.02$ at symmetric average field $\overline{\psi}=0$ (see simulation details in SM~\cite{supmat} Sec. IV and Fig.~S4 in SM~\cite{supmat} for the phase diagram). 
Fig.~\ref{MainFigFSS}(a) displays the structure factor $S(q)$ near the LG critical point, confirming two key features: (i) the predicted  $q^2$ scaling of hyperuniformity below the critical temperature, and (ii) {finite} fluctuation behavior $S(q\to 0)\sim {const.}$ precisely at the critical point.
The spatial correlation functions $C(x)$ in Fig.~\ref{MainFigFSS}(b) demonstrate $\delta$-function characteristics near criticality, in agreement with Eq.~(\ref{correlation_func}) {and spinner systems in Fig.~\ref{MainFigCor}}. Here, quantities marked with a tilde denote the corresponding dimensionless quantities.
The response function $\chi(x)$, shown in the inset of Fig.~\ref{MainFigFSS}(b), is in stark contrast to the correlation function $C(x)$, indicating a violation of the conventional FDR at a fundamental level~\cite{Young2020}. We also show the log-log plot of $\chi(x)$ in Fig.~S5 in SM~\cite{supmat} which gives $\eta'=0.26(2)$, while $\eta$ is unattainable from $C(x)$. 
In Fig.~\ref{MainFigGFDR}(b, c), we further verify the generalized FDR, Eq.~(\ref{GFDR}) and Eq.~(\ref{GFDRstatic}), \add{by computing dynamic structure factor and response function in the momentum-frequency space(SM~\cite{supmat} Sec. III and Sec. IV)} . These results confirm that the effective temperature in the system scales as $k_BT_{\rm eff}\sim q^2$.

\begin{table*}[!htb] 
\begin{tabular}{c c c c c c c c c c c}
\hline \\[-2.0ex]
~~~~~&~~~~~$\eta$~~~~~&~~~~~$\eta'$~~~~~
& ~~~~~$\gamma$~~~~~ &~~~~~$\gamma'$~~~~~
& ~~~~~$\nu$~~~~~
& ~~~~~$\beta$~~~~~ 
& ~~~~$\delta$ ~~~~~&~~~~~$\nu_{\parallel}$~~~~  \\[1.5ex]
\hline 
\\[-1.5ex]

{MF Ising}~~  &0 &0
& 1 & 1
& $\dfrac{1}{2}$
& $\dfrac{1}{2}$ 
& 3  & $2$   \\ [1.5ex]

{2D Ising}    & $\dfrac{1}{4}$ & $\dfrac{1}{4}$
& $\dfrac{7}{4}$  & $\dfrac{7}{4}$
&1
& $\dfrac{1}{8}$ 
& 15
&  $\dfrac{15}{4}$  \\ [1.5ex]

{2D HU fluids}  &0.0(1) & 0.0(1)*
& 0.99(2) & 0.99(2)
& 0.50(2) 
& 0.51(2) 
& 2.98(6)  
&  2.0(1) \\ [1.5ex]

\hline \\

\end{tabular}
\caption{ Comparison of critical exponents of 2D HU fluids with the Ising universality class~\cite{Pelissetto2002}. 
\add{Note that the critical exponents $\gamma$ and $\gamma'$, as well as $\eta$ and $\eta'$, are measured from the divergence of compressibility and density fluctuations, respectively. Different from equilibrium cases, they are not necessarily the same in HU fluids.}
*This result is obtained from the compressibility in Fig.~\ref{MainFigFSS}(f), while by directly measuring the response function (Fig.~S5 in SM~\cite{supmat}), we obtain $\eta'=0.26(2)$.} \label{Tab_1}
\end{table*}

For the LG phase transition, the order parameter is  the density difference between liquid and gas phases, $\Delta \psi_{lg} = \psi_l - \psi_g$. Fig.~\ref{MainFigFSS}(c) shows $\Delta \psi_{lg}$ as a function of the distance to the critical point $\tau$. Through finite-size scaling
\begin{equation}
\Delta \psi_{lg}(\tau,L) = L^{-\beta/\nu}\mathcal{G}(\tau L^{1/\nu}),
\end{equation}
we extract critical exponents $\beta=0.51(2)$ and $\nu=0.50(2)$, consistent with mean-field values $\beta_{\rm MF}=1/2$ and $\nu_{\rm MF}=1/2$. Fig.~\ref{MainFigFSS}(d) demonstrates the response of $\Delta \psi_{lg} $ to an external field $h$ near the critical point, which obeys the finite-size scaling 
\begin{equation}
\Delta \psi_{lg}(h,L) =L^{-\beta/\nu}{\Psi}(h L^{\beta\delta/\nu})
\end{equation}
from which we extract critical exponents $\delta=2.98(6)$  ($\delta_{\rm MF}=3$).
Figs.~\ref{MainFigFSS}(e-f) compare density fluctuations $\chi_\rho$ and compressibility $\chi_c$ near criticality. We observe contrasting scaling behaviors: while $\chi_c$ increases with system size $L$, $\chi_\rho$ decreases with $L$. 
This non-conventional critical phenomenon is described by our proposed scaling in Eqs.~(\ref{compressibility_finite_size}-\ref{S0_finite_size}), from which we determine $\gamma=0.99(2),~\gamma'=0.99(2)$ ($\gamma_{\rm MF}=1$). Based on scaling relations $\gamma=\nu(2-\eta)$ and $\gamma'=\nu(2-\eta')$, one obtains $\eta=0.0(1),~\eta'=0.0(1)$.
Fig.~\ref{MainFigFSS}(g) displays Binder cumulant~\cite{Maggi2021}
\begin{equation}
U_4 = \langle\Delta\psi^2\rangle^2/\langle\Delta\psi^4\rangle.
\end{equation}
Distinct from the Ising universality class, the $U_4$ curves for different system sizes converge rather than intersect at value $U_4^*=1/3$ indicating Gaussian fluctuation of HU fluids at criticality~\cite{binder1992monte}.
Note that the critical Gaussian fluctuation is not due to the system at the upper critical dimension. In fact, non-Gaussian fluctuations are an important characteristic of the Ising universality class even at mean-field (mean-field value is $U_4^*= 0.456947$)~\cite{lv2021finite,luijten1997interaction}.  
This non-Gaussian behavior is a result of the vanishing quadratic energy penalty for fluctuations at $q=0$, thus leaving the nonlinear coupling terms to become dangerously irrelevant~\cite{CriDyn}.
For HU fluids, since the effective temperature vanishes as $q \rightarrow 0$, this nonlinear coupling becomes `safely irrelevant' in this case.

\begin{figure*}[!bhtp]
	\resizebox{180mm}{!}{\includegraphics[trim=0.0in 0.0in 0.0in 0.0in]{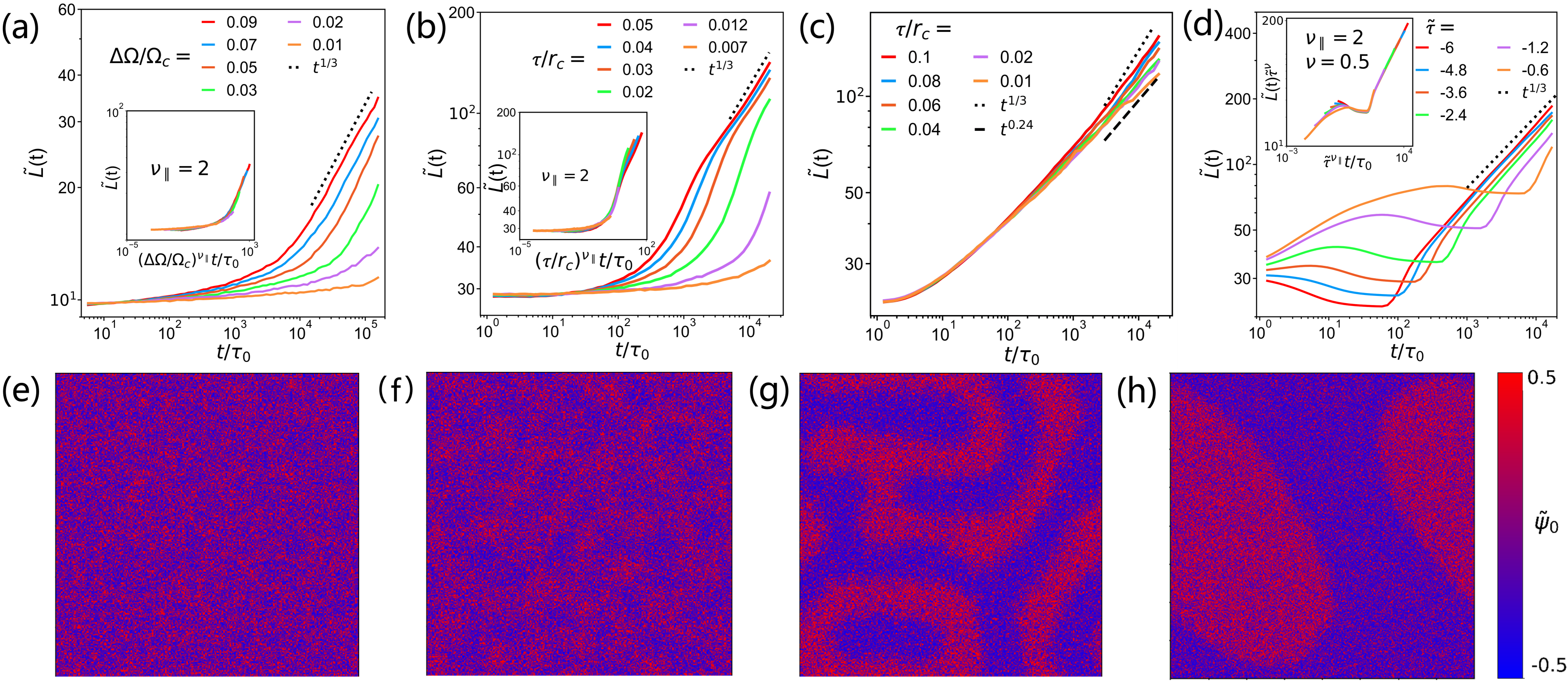} }
\caption{The spinodal decomposition and coarsening process of  (a) {2D} active spinner fluids ($N=40,000$), (b) 2D stochastic field, (c) 2D classical Model B  and (d)  2D  Cahn-Hilliard equation at different distances ($\tau/r_c$ or $\Delta \Omega/\Omega_c$) from the critical point. Snapshots of 2D stochastic field with $\tau/r_c=-0.05$ at (e) $ \tau_0$, (f) $1.5\times 10^{2}\tau_0$, (g)  $9\times 10^{2}\tau_0$ and (h)  $4\times 10^{5}\tau_0$. The system size of field simulation is $800\times800$. {Corresponding movies are also provided in SM.}}   \label{MainFigDyna}
\end{figure*}

Fig.~\ref{MainFigFSS}(h) further shows the energy fluctuation of HU fluids near the critical point, which obeys an anomalous finite-size scaling different from equilibrium fluids 
\begin{equation}
c_v(\tau,L) = f(\tau) + c_0 L^{-\lambda}+ c_1 \tau L^{1/\nu - \lambda}
\end{equation}
with $\lambda=2.11,~c_0=24.0,~c_1=13.0$ {(see Appendix E Sec. 3)}. 
{This is distinct from the logarithmic divergence of the 4D Ising model $c_v(\tau,L)=\ln^{1/3} L\cdot K(\tau L^2 \ln^{1/6} L)$~\cite{luijten1997interaction,aktekin2001finite,li2024logarithmic}, as well as the Ising model above 4D $c_v(\tau,L)=K(\tau L^2 )$~\cite{luijten1997interaction,kalay2007finite}. }

Finally, we summarize all critical exponents obtained from the field simulation in Table~\ref{Tab_1}, to compare it with the mean-field value and the 2D Ising model.  
Note that the $\eta'$ measured from the response function ({Fig.~S5 in SM~\cite{supmat}}) does not match that obtained from compressibility $\chi_c$. 
This inconsistency may be due to the limited accuracy in measuring the response function close to the critical point, or for other mechanism not considered by our current theoretical framework. 
{Moreover, while logarithmic corrections are generally expected for systems at the upper critical dimension, our stochastic field simulations lack the precision needed to resolve them.}

%The origin of this inconsistency would be our ongoing topic.

%At high temperatures the energy fluctuations is independent of  system size, whereas at low temperatures, the energy fluctuations obey the finite size scaling  $c_v \sim L^{-0.063}$ suggesting a hyperuniform phase interface distribution.

\section{Spinodal Decomposition and Coarsening Dynamics}
In this section, we investigate the dynamics of LG phase transitions of HU fluids by analysing the characteristic length scale $L(t)$ of inhomogeneities, defined as~\cite{clewett2012emergent,Shi2020}:
\begin{eqnarray}
	{L}(t)= \left\langle  \frac{\int  q S(q, t) d q}{ 2 \pi \int S(q, t) d q} \right\rangle ^{-1}
\end{eqnarray}
where $S(q, t)$ denotes the time-dependent structure factor.  Fig.~\ref{MainFigDyna}(a) shows $L(t)$ the evolution of initially homogeneous active spinner systems quenched into phase separation at the critical density (by increasing $\Omega$ over critical $\Omega_c $). Three phenomena emerge: (i) a pre-coarsening stage in $L(t)$ when $t<t_w$, (ii) subsequent coarsening with $L(t) \sim t^{1/3}$ when $t>t_w$, and (iii) critical divergence of waiting time $t_w \sim (\Omega-\Omega_c)^{-\nu_{\parallel}}$ with $\nu_{\parallel}=2.0(1)$. 
Stochastic field simulations of Model B-like dynamics of HU fluids (Fig.~\ref{MainFigDyna}(b)) reproduce this pre-coarsening stage but reveal an additional fast intermediate regime between pre-coarsening and coarsening~\cite{clewett2012emergent}.
The difference between spinner systems and field simulations is probably due to the odd viscosity in spinner systems, which can affect the interfacial dynamics~\cite{soni2019odd}. 
For comparison, we show the coarsening dynamics of the \add{2D and 4D}  classical Model B in Fig.~\ref{MainFigDyna}(c) and Fig.~S6 in SM~\cite{supmat}, respectively.   \add{The different coarsening dynamics can also be identified in Movies S5-S6 of SM~\cite{supmat}.} We find the absence of the waiting time and a slower coarsening scaling as the system is quenched close to the LG critical point. Departing from the critical point, the coarsening dynamics of the above three systems all follow  $L(t) \sim t^{1/3}$~\cite{livi2017nonequilibrium}.

The existence of a waiting time before coarsening resembles the dynamics of the noiseless Cahn-Hilliard equation, where the spinodal decomposition with the characteristic time $t_w$ is observed before the coarsening dynamics. 
This characteristic spinodal decomposition time can be estimated in a mean-field theory with linear perturbations about the stable state $\psi = \psi_0 + \delta\psi$ in Eq.~(\ref{CMC-modelB-s}), which gives the growth rate of the perturbation of the wave vector $k$ below the critical temperature ($r<0$):
\begin{eqnarray}
    \omega(k) = - \left(r+\frac{1}{2}u\psi^2_0\right)k^2- k^4.
\end{eqnarray}
At the critical density $\psi_0=\psi_c=0$, the growth rate reaches its maximum value at the most unstable wavevector $k_{max}=\sqrt{\frac{-r}{2}}$~\cite{kostorz2001phase}, which gives the characteristic length scale
\begin{eqnarray} \label{coarseninglength}
L_c = \frac{2\pi}{k_{max}}=2\pi\sqrt{\frac{2}{-r}}\sim r^{-1/2}
\end{eqnarray}
The time scale for growth of the most unstable mode is
\begin{eqnarray}\label{waitingTime}
    t_w=  \frac{2\pi}{\omega(k_{max})} =   \frac{8\pi}{r^2} \sim r^{-2}
\end{eqnarray}
Thus, $L_c$ and $t_w$ diverge at the critical point {according to the scaling law $L_c \sim  \tau^{-\nu}$ and $t_w \sim  \tau^{-\nu_{\parallel}}$, respectively, with mean-field critical exponents $\nu=1/2$ and $\nu_{\parallel}=2$.}  The data collapse of the Cahn-Hilliard equation $\Tilde{L}(t)$ in Fig.~\ref{MainFigDyna}(d) confirms these scaling laws. 
The divergent waiting time is absent in equilibrium fluids (see Fig.~\ref{MainFigDyna}(c) and Fig.~S6 in SM~\cite{supmat}), because divergent critical fluctuations smear the spinodal decomposition. 
For HU fluids, however, long-wavelength density fluctuations are significantly suppressed. Thus, the decomposition time before coarsening can still be  observed near the critical point. Nevertheless, the decomposition length does
not show divergence in HU fluids as shown in Fig.~\ref{MainFigDyna}(a-b), {indicating that fluctuations still play an important role in the length scale selection}. In all, the spinodal decomposition in HU fluids is fundamentally different from either the conventional scenario of the Model B or that of athermal phase separation.

\section{Conclusion and Discussion}

In this work, by combining field theory and molecular dynamics simulations, we systematically study the LG phase separation of non-equilibrium HU fluids with additional center-of-mass  conservation. As a concrete realization, we focus on active spinner fluids whose LG phase separation can be induced by dissipative collisions. A hydrodynamic theory is proposed to connect the microscopic spinner model to the field theory of HU fluids. With both analytical and numerical evidence, we establish that HU fluids are unusually calm yet extremely susceptible at LG criticality, distinct from the Ising universality class in several aspects: i) upper critical dimension decreases from 4 to 2; ii) Gaussian type critical fluctuation; iii) short-range pair correlation at criticality; iv) non-divergent \add{density/energy} fluctuations and their non-conventional finite-size scaling at criticality; v) non-conventional spinodal decomposition dynamics. The origin of these non-conventional critical behaviors lies in the special non-equilibrium nature of the system which fundamentally violates the conventional FDR, but respects a generalized FDR with $q$-dependent effective temperature. 
Note that hyperuniform fluids and corresponding non-conventional LG criticality are sensitive to thermal noise that violates center-of-mass  conservation. Thus, we expect that our predictions to be experimentally tested in macroscopic systems such as active spinner systems~\cite{Scholz2018a,li2023inertial,wang2024robo,wang2025active}, vibration driven granular gases~\cite{clewett2012emergent,clewett2016minimization,herminghaus2017,maire2024interplay,maire2025dynamical}, pulsating robotic systems~\cite{li2019particle}, or low temperature microscopic systems such as light-driven atomic systems~\cite{chu1985three,grant1998dielectric,villeneuve2000forced}. In these systems, the effect of non-conserved noise can be effectively reduced by increasing either the experimental accuracy or the strength of center-of-mass conserved activity. Studies in this direction would open new routes to realize and probe non-conventional critical phenomena, as well as fractonic physics, in soft matter systems~\cite{yuan2020fractonic,gromov2020fracton,guardado2020subdiffusion,
glorioso2022breakdown,han2024scaling,nandkishore2019fractons}.

\vspace{1em}
\noindent\textit{Note added--}
{Recently, Ref.~\cite{maire2025hyperuniform} appeared, which independently proposes a similar HU Model B to study interface fluctuations of phase separated HU fluids.}

\section*{Acknowledgments}

The authors acknowledge fruitful discussions with Ran Ni, Xia-qing Shi, Leiming Chen, Rapha{\"e}l Maire, Xiaosong Chen, Bing Miao, Youjin Deng, Qing Yang and Yu Duan.  This work is supported by  the National Natural Science Foundation of China (No.~12347102, 12275127), the Natural Science Foundation of Jiangsu Province (No. BK20250058, No. BK20233001), \add{Program for Innovative Talents and Entrepreneur in Jiangsu,}  the Fundamental Research Funds for the Central Universities (0204-14380249, KG202501), the Fundamental and Interdisciplinary Disciplines Breakthrough Plan of the Ministry of Education of China (No. JYB2025XDXM502), Quantum Science and Technology-National Science and Technology Major Project (No. 2024ZD0300101). The simulations were performed on the High-Performance Computing Center of Collaborative Innovation Center of Advanced Microstructures, the High-Performance Computing Center (HPCC) of Nanjing University.

\add{
\section*{Data Availability}

The data and codes that support the findings of this article are openly available~\cite{open}.
}

\setcounter{figure}{0}
\renewcommand{\thefigure}{A\arabic{figure}}
\setcounter{equation}{0}
\renewcommand{\theequation}{A\arabic{equation}}

\setcounter{secnumdepth}{2}
\setcounter{subsection}{0}

\setcounter{table}{0}
\renewcommand{\thetable}{A\Roman{table}}
\renewcommand{\thesubsection}{\arabic{subsection}}

~~~\\

~~~\\

\section*{Appendix A: Generalized fluctuation-dissipation relation}

HU fluids satisfy the generalized fluctuation-dissipation relation Eq.~(\ref{GFDR}).
To prove this, we first perform a Fourier transformation of the field equation Eq.~(\ref{CMC-modelB-minimal1}) with the external field $h$, i.e.,
\begin{equation}\label{CMC-modelB-minimal3}
    \frac{\partial \psi(\mathbf{q},t)}{\partial t}
    =
    -q^2\left[
    \add{\Tilde{\mathcal{J}}[\psi](\mathbf{q},t)}
    -
    h(\mathbf{q},t)
    \right] + \zeta_{\psi}(\mathbf{q},t)
\end{equation}
\add{where $\psi(\mathbf{q},t)=\int d^dx~ e^{-i\mathbf{q}\cdot\mathbf{x}}\psi(\mathbf{x},t)$, $\zeta_{\psi}(\mathbf{q},t)=\int d^dx~ e^{-i\mathbf{q}\cdot\mathbf{x}}$ $\nabla^2\eta_{\psi}(\mathbf{x},t)$ and $\Tilde{\mathcal{J}}[\psi](\mathbf{q},t)=\int d^dx~ e^{-i\mathbf{q}\cdot\mathbf{x}} \mathcal{J}[\psi](\mathbf{x},t)$.}
To derive Eq.~(\ref{GFDR}), we first obtain the Onsager-Machlup functional of Eq.~(\ref{CMC-modelB-minimal3})~\cite{CriDyn}
\begin{widetext}
\begin{align}
    \mathcal{G}_{h}[\psi]
    =&
    \frac{1}{4}\int \add{\frac{d^dq}{(2\pi)^d} }\int dt~ \frac{1}{Dq^{4}}
    \left[
    \frac{\partial\psi(-\mathbf{q},t)}{\partial t} + q^2\left( \Tilde{\mathcal{J}}[\psi](-\mathbf{q},t) - h(-\mathbf{q},t) \right)
    \right]
    \left[
    \frac{\partial\psi(\mathbf{q},t)}{\partial t} + q^2\left( \Tilde{\mathcal{J}}[\psi](\mathbf{q},t) - h(\mathbf{q},t) \right)
    \right]\nonumber\\
    =&
    \mathcal{G}_{h=0}[\psi]
    -
    \frac{1}{2} \int \add{\frac{d^dq}{(2\pi)^d}}\int dt~\frac{1}{Dq^{2}}
    \left[
        \frac{\partial \psi(-\mathbf{q},t)}{\partial t}
        +
        q^2 \Tilde{\mathcal{J}}[\psi](-\mathbf{q},t)
    \right] h(\mathbf{q},t) + O(h^2)
\end{align}
\end{widetext}
Using the statistical weight $\mathcal{P}[\psi]= e^{-\mathcal{G}_h}$, the dynamical susceptibility is then given by
\begin{align}
    &\chi(\mathbf{q},t-t')
    =
    \left.\frac{\add{\delta}\langle \psi(\mathbf{q},t) \rangle}{\add{\delta} h(\mathbf{q},t')}\right|_{h=0}
    =
    \left.\frac{\add{\delta} \int \mathcal{D}[\psi]~e^{-\mathcal{G}_h} \psi(\mathbf{q},t) }{\add{\delta} h(\mathbf{q},t')}\right|_{h=0}\nonumber\\
    &=
    \frac{1}{2\add{(2\pi)^d}Dq^{2}}
    \left\langle   
        \psi(\mathbf{q},t)
        \left[ 
            \frac{\partial \psi(-\mathbf{q},t')}{\partial t'}
            +
            q^2 \Tilde{\mathcal{J}}[\psi](-\mathbf{q},t')
        \right]
    \right\rangle \label{qtSusceptibility}
\end{align}
Causality demands that $\chi(\mathbf{q},t-t')=0$ for $t<t'$, hence
\begin{equation}\label{Causality}
    t<t':~~
    \left\langle   
        q^2 \psi(\mathbf{q},t)
        \Tilde{\mathcal{J}}[{\psi}](-\mathbf{q},t')
    \right\rangle
    =
    -
    \left\langle   
        \psi(\mathbf{q},t) 
            \frac{\partial \psi(-\mathbf{q},t')}{\partial t'}
    \right\rangle
\end{equation}
% Next, we consider another solution to Eq.~(\ref{fieldEquFourier}), a time inversion process (apply $\psi(\mathbf{q},t)\to\psi(\mathbf{q},-t)$ of Eq.~(\ref{fieldEquFourier}). Note that the system is not required to be equilibrium, so there is no need for the probabilities of the two processes to be equal, then we can obtain an equation using the same way as above 
% \begin{eqnarray}\label{CausalityInverse}
%     t<t':~~
%     \left\langle   
%         q^2 \psi(\mathbf{q},-t)
%         \Tilde{G}(-\mathbf{q},-t')
%     \right\rangle
%     =
%     -
%     \left\langle   
%         \psi(\mathbf{q},-t) 
%             \frac{\partial \psi(-\mathbf{q},-t')}{\partial t'}
%     \right\rangle
% \end{eqnarray}
% Relabeling $t\to -t,~t'\to -t'$, then we obtain
% \begin{eqnarray}\label{Causality2}
%     t>t':~~
%     \left\langle   
%         q^2 \psi(\mathbf{q},t)
%         \Tilde{G}(-\mathbf{q},t')
%     \right\rangle
%     =
%     \left\langle   
%         \psi(\mathbf{q},t) 
%             \frac{\partial \psi(-\mathbf{q},t')}{\partial t'}
%     \right\rangle
% \end{eqnarray}
Next, we consider another solution to Eq.~(\ref{CMC-modelB-minimal1}), a time reversed process $\psi'(x,\delta t)=\psi(x,t'-\delta t)$~\cite{CriDyn}. Then, following the same procedure as above, we obtain
\begin{align}
    \delta t<\delta t':~~
    &\left\langle   
        q^2 \psi'(\mathbf{q},\delta t)
        \Tilde{\mathcal{J}}[{\psi}'](-\mathbf{q},\delta t')
    \right\rangle \nonumber\\
    &=
    -
    \left\langle   
        \psi'(\mathbf{q},\delta t) 
            \frac{\partial \psi'(-\mathbf{q},\delta t')}{\partial \delta t'}
    \right\rangle \label{CausalityInverse}
\end{align}
i.e.,
\begin{align}
    \delta t<\delta t':~~
    &\left\langle   
        q^2 \psi(\mathbf{q},t'-\delta t)
        \Tilde{\mathcal{J}}[{\psi}](-\mathbf{q},t'-\delta t')
    \right\rangle \nonumber\\
    &=
    -
    \left\langle   
        \psi(\mathbf{q},t'-\delta t) 
            \frac{\partial \psi(-\mathbf{q},t'-\delta t')}{\partial \delta t'}
    \right\rangle
\end{align}
Let $\Tilde{\delta t}=\delta t-t',~\Tilde{\delta t}'=\delta t'-t'$, then we have
\begin{align}
    \Tilde{\delta t}<\Tilde{\delta t}':~~
    &\left\langle   
        q^2 \psi(\mathbf{q},-\Tilde{\delta t})
       \Tilde{\mathcal{J}}[{\psi}](-\mathbf{q},-\Tilde{\delta t}')
    \right\rangle \nonumber\\
    &=
    -
    \left\langle   
        \psi(\mathbf{q},-\Tilde{\delta t}) 
            \frac{\partial \psi(-\mathbf{q},-\Tilde{\delta t}')}{\partial \Tilde{\delta t}'}
    \right\rangle
\end{align}
Relabeling $\Tilde{\delta t}\to -t,~\Tilde{\delta t}'\to -t'$, then we obtain
\begin{equation}\label{Causality2}
    t > t':~~
    \left\langle   
        q^2 \psi(\mathbf{q},t)
        \Tilde{\mathcal{J}}[{\psi}](-\mathbf{q},t')
    \right\rangle
    =
    \left\langle   
        \psi(\mathbf{q},t) 
            \frac{\partial \psi(-\mathbf{q},t')}{\partial t'}
    \right\rangle
\end{equation}

Inserting Eq.~(\ref{Causality}) and Eq.~(\ref{Causality2}) into Eq.~(\ref{qtSusceptibility}) yields the dynamical generalized fluctuation-dissipation relation
\begin{align}
    \chi(\mathbf{q},t-t')
    &=
    -\frac{1}{\add{(2\pi)^d} Dq^{2}}\Theta(t-t')\frac{\partial}{\partial t}\langle \psi(\mathbf{q},t)\psi(-\mathbf{q},t') \rangle \nonumber\\
    &=
     -\frac{1}{Dq^{2}}\Theta(t-t')\frac{\partial}{\partial t}C(q,t-t')
\end{align}\label{GFDT}
where $\Theta(t-t')$ is the Heaviside function.
We can define a scale-dependent effective temperature $k_B T_{\text{eff}}(q) = Dq^{2}$. Upon Fourier transform~\cite{CriDyn}, we obtain the dynamical generalized fluctuation-dissipation relation in frequency space, i.e., Eq.~(\ref{GFDR}),
\begin{eqnarray}
    \mathrm{Im}~ \chi(q,\omega) = \frac{\omega }{2Dq^2}C(q,\omega).
\end{eqnarray}
According to the Kramers-Kronig relation, we have
\begin{align}
    \mathrm{Re}~ \chi(q,\omega)
    &=
    \frac{1}{\pi} \add{\mathcal{P} } \int \frac{\mathrm{Im}~ \chi(q,\omega')}{\omega'-\omega}d\omega' \nonumber\\
    &=
    \frac{1}{2\pi Dq^2}
    \add{\mathcal{P} } \int \frac{\omega'C(q,\omega')}{\omega'-\omega}d\omega'
\end{align}
\add{where $\mathcal{P}$ denotes the Cauchy principal value.}
Considering $\chi(q,\omega=0)=\mathrm{Re}~\chi(q,\omega=0)$, we obtain the static generalized fluctuation-dissipation relation
\begin{align}
    \chi(q) 
    &\equiv 
    \chi(q,\omega=0) \nonumber\\
    &=
    \frac{1}{ Dq^2}
    \int C(q,\omega')\frac{d\omega'}{2\pi}
    \equiv \frac{S(q)}{Dq^2}=\frac{S(q)}{k_BT_{\text{eff}}(q)}
\end{align}

%\clearpage

\section*{Appendix B: Derivation of Hydrodynamic Theory of Active Spinners}

Here we present the detailed derivation of Eqs.~(\ref{contiEq}-\ref{kineticEqn2}).
Eq.~(\ref{contiEq}) is the continuity equation for  the conserved scalar density field $\rho(\mathbf{x},t)$. Eq.~(\ref{NSEq}) is the momentum equation for fluids with odd viscosity~\cite{fruchart2023odd}.
The linear damping term $- \gamma_t  \rho \mathbf{u}$ describes the frictional force between spinners and substrate. $\nabla \cdot \boldsymbol{\sigma}_{ \mathbf{u}}$ is the momentum-conserved white noise. Eqs.~(\ref{kineticEqn1}-\ref{kineticEqn2}) are derived from the energy equations for the translational and rotational degrees of freedom
\begin{align}
\frac{\mathcal{D}T_t}{\mathcal{D}t} =& D_t\nabla^2 T_t  - A_{\text{friction}} + A_{\text{gain}} - A_{\text{loss}}  + \eta_t(\mathbf{x},t)  \\
\frac{ \mathcal{D} T_r}{\mathcal{D} t}  =& D_r\nabla^2 T_r  - B_{\text{friction}}
   + B_{\text{gain}}   - B_{\text{loss}} + B_{\text{drive}} + \eta_r(\mathbf{x},t), 
\end{align}
where $D_t\nabla^2 T_t,D_r\nabla^2 T_r$ are diffusion terms, reflecting how kinetic energy diffuses from high-energy regions to low-energy regions through collisions. 
$\eta_t,~\eta_r$ are Gaussian white noises.    $A_{\text{friction}}$ and $B_{\text{friction}}$ are the translational and rotational energy damping terms, representing the power dissipated by the frictional force $-\gamma_t v$ and rotational frictional torque $-\gamma_r \omega$, i.e.,
\begin{eqnarray}
A_{\text{friction}} &=& -\gamma_t v\cdot v = - \frac{2\gamma_t }{m} T_t  \\
B_{\text{friction}} &=& -\gamma_r \omega \cdot \omega = - \frac{2\gamma_r }{I}T_r.
\end{eqnarray}
$B_{\text{drive}}$ is the energy driving term, representing the power injected by the external torque into the spinner rotation, i.e.,
\begin{eqnarray}
B_{\text{drive}}=\Omega \omega=\Omega\cdot\sqrt{\frac{2}{I}} T_r^{1/2}=\frac{2\gamma_r }{I}T_{ss}^{1/2}T_r^{1/2}
\end{eqnarray}
where  $T_{ss} = I\Omega^2/(2\gamma_r^2)$. Note that $B_\text{drive}=B_\text{friction}$ when $T_r=T_{ss}$. Thus, $T_{ss}$ is the steady rotational kinetic energy of an isolated spinner.
$A_{\text{gain/loss}}$ and $ B_{\text{gain/loss}}$ terms account for the energy gain and loss due to inter-spinner collisions, which allow energy exchange between translational and rotational kinetic energies. 
For a system under thermal equilibrium, this gives rise to the equipartition theorem. According to the molecular kinetic theory,  these four terms are proportional to the average collision frequency $\bar{Z}$. As a first order approximation,  $A_{\text{gain/loss}}$ and $ B_{\text{gain/loss}}$ can be written as~\cite{luding1998homogeneous}
\begin{eqnarray}
A_{\text{gain}}=s_1 \Bar{Z} T_r,~~ A_{\text{loss}} =s_2 \Bar{Z} T_t \\
B_{\text{gain}}=s_3 \Bar{Z} T_t,~~ B_{\text{loss}} =s_4 \Bar{Z} T_r.
\end{eqnarray}
Here $s_1, s_2, s_3, s_4$ are the collision coefficients  to account for energy exchange and dissipation during collision~\cite{luding1998homogeneous}. For a hard sphere gas, the collision frequency is $\bar{Z}_0= \bar{v}/{\lambda}$, where $\bar{v}=\sqrt{ {2T_t}/{m} }$ and $\lambda = 1/(\sqrt{2 \pi}\rho\sigma )$ are the mean speed and the mean free path, respectively~\cite{cutchis1977enskog,visco2008collisional}.  In our case, however, the next collision may not occur due to the damping effect on the translational velocity. 
The strength of damping can be reflected by the average distance that an isolated spinner can travel, $l_d \sim  A  \sqrt{m T_t}/\gamma_t$~\cite{Lei2019h}.  In classical molecular theory, the survival probability of particles without collision after travelling distance $l_d$ is $e^{-l_d/\lambda}$, which can be used to estimate the probability that the next collision is suppressed by damping. Thus, $\bar{Z}$ should be modified by the next-collision survival probability $(1-e^{-l_d/\lambda})$, i.e.,
\begin{eqnarray}
\bar{Z}=  (1-e^{-l_d/\lambda})\bar{Z}_0= (1-e^{-l_d/\lambda})\bar{v}/{\lambda}.
\end{eqnarray}
Based on the simulation data of spinners in Fig.~\ref{MainFigHydro}(a), we obtain $A\approx 4$ in the coefficient in $l_d$.  Then, we obtain the complete form of Eqs.~(\ref{kineticEqn1}-\ref{kineticEqn2}).

\section*{Appendix C: Derivation of Model-B-like Equation and Effective Free Energy}

For the momentum equation Eq. (\ref{NSEq}),  the drag term $- {\gamma_t} \rho \mathbf{u}$ damps the velocity field  $\mathbf{u}(\mathbf{r},t)$, so the inertia term $\frac{ \mathcal{D} \mathbf{u}}{\mathcal{D} t}=\left( \frac{\partial }{\partial t} + \mathbf{u}\cdot\nabla \right)\mathbf{u}$ can be neglected in the long-wavelength limit. Thus we have 
\begin{eqnarray}\label{NS}
    0 = 
    -\nabla P + {\eta^*}\nabla^2 \mathbf{u} 
    + {\zeta}_0\nabla(\nabla\cdot \mathbf{u})
    - {\gamma_t} \rho \mathbf{u}
    + \nabla\cdot\boldsymbol{\sigma}_{\mathbf{u}}.~~~~
\end{eqnarray}
Assuming small perturbations around the steady state $\rho(\mathbf{r},t)=\rho_0,~T_t(\mathbf{r},t)=T_{t0},~P(\mathbf{r},t)=P_0,~\mathbf{u}(\mathbf{r},t)=0$, we have
\begin{eqnarray}
   &&\delta \rho = \rho(\mathbf{r},t) -  \rho_0,~   \delta T =T_t(\mathbf{r},t)-T_{t0} ,\nonumber\\
   &&\delta P = P(\mathbf{r},t)-P_0,~ \delta \mathbf{u} = \mathbf{u}(\mathbf{r},t).  \label{pert}
\end{eqnarray}
Then Eqs.~(\ref{contiEq}-\ref{NSEq}) reduce to a set of linear equations:
\begin{align}
    &\frac{\partial \delta \rho}{\partial t} + \rho_0 \nabla\cdot\left( \delta \mathbf{u} \right) = 0\\
    &0=-\nabla\delta P + \eta^* \nabla^2\delta \mathbf{u} 
    + {\zeta}_0\nabla(\nabla\cdot \delta \mathbf{u})
    - {\gamma_t} \rho_0 \delta \mathbf{u} + \nabla\cdot\boldsymbol{\sigma}_{\mathbf{u}}.
\end{align}
Under Fourier transformation, e.g.,  $\widetilde{ \delta \rho } = \int dt\int d^dx\ e^{i\omega t}e^{-i\mathbf{q}\cdot\mathbf{x}}\delta \rho$ and using Helmholtz decomposition $\widetilde{\delta\mathbf{u}}=\widetilde{\delta\mathbf{u}}_{\parallel}+\widetilde{\delta\mathbf{u}}_{\perp}$~\cite{gross2011modelling}, we can obtain
\begin{align}
    &-i\omega \widetilde{\delta \rho} = i\rho_0\mathbf{q}\cdot  \widetilde{\delta\mathbf{u}}_{\parallel}\label{continuity_k} \\
    &i\mathbf{q}\widetilde{\delta P} 
    - (\eta^*+\zeta_0) q^2 \widetilde{\delta \mathbf{u}}_{\parallel} 
    - {\gamma_t} \rho_0 \widetilde{\delta \mathbf{u}_{\parallel}} - i \mathbf{q} \widetilde{\sigma}_{\parallel, u}=0\label{NS_k}\\
    &i\mathbf{q}\widetilde{\delta P} 
    - \eta^* q^2 \widetilde{\delta \mathbf{u}}_{\perp}
    -{\gamma_t} \rho_0 \widetilde{\delta \mathbf{u}_{\perp}}
    - i \mathbf{q} \widetilde{\sigma}_{\perp, u}=0.
\end{align}
Generally, the pressure field $P$ can be expanded in terms of the local fields and the local field derivatives, i.e.,
\begin{equation}\label{delta_P}
    \delta P(\rho,T_t,\nabla \rho, \nabla T_t, \nabla^2 \rho, \cdots)
    =
    \frac{1}{\Bar{\chi}_T} \delta\rho + \beta_V\delta T 
    -
    K_{\rho}\nabla^2\delta\rho +\cdots.
\end{equation}
Note that the first order derivative terms are zero, as required by the reflection and rotational symmetries of the pressure~\cite{Cahn1958}.  Combining Eq.~(\ref{delta_P}) and Eq.~(\ref{NS_k}), we  obtain
\begin{eqnarray}\label{delta_u}
    \mathbf{A}\cdot\widetilde{\delta \mathbf{u}}_{\parallel}={ i\mathbf{q}\left[ 
        \left(\frac{1}{\Bar{\chi}_T}+K_{\rho}q^2\right)\widetilde{\delta \rho}
        +
        \beta_V\widetilde{\delta T}
        -  \widetilde{\sigma}_{\parallel, u}
    \right] }
\end{eqnarray}
where
\begin{eqnarray}
    \mathbf{A}
    =
    \begin{pmatrix}
        (\eta_0+\zeta_0)q^2+\rho_0 \gamma_t & \eta^o q^2\\
        -\eta^o q^2 & (\eta_0+\zeta_0)q^2+\rho_0 \gamma_t.
    \end{pmatrix}
\end{eqnarray}
Combining Eq.~(\ref{delta_u}) and Eq.~(\ref{continuity_k}), we  obtain 
\begin{eqnarray}\label{delta_phi}
    -i\omega \widetilde{\delta \rho} 
    &&=
    -   { \rho_0\left[ 
        \left(\frac{1}{\Bar{\chi}_T}+K_{\rho}q^2\right)\widetilde{\delta \rho}
        +
        \beta_V\widetilde{\delta T}
        -  \widetilde{\sigma}_{\parallel, u}
    \right] }
    \mathbf{q}\cdot\mathbf{A}^{-1}\cdot\mathbf{q}
    \nonumber \\
    &&=
    -\rho_0 q^2 \dfrac{ 
        \left(\dfrac{1}    {\Bar{\chi}_T}+K_{\rho}q^2\right)\widetilde{\delta \rho}
        +
        \beta_V\widetilde{\delta T}
        -  \widetilde{\sigma}_{\parallel, u}
    }
    {
        (\eta_0+\zeta_0)q^2+\rho_0 \gamma_t + \dfrac{\eta^{o2}q^4}{(\eta_0+\zeta_0)q^2+\rho_0 \gamma_t}
    }
    \nonumber\\
    &&\approx 
    -\frac{ q^2}{\gamma_t} 
    \left[ 
        \left(\frac{1}{\Bar{\chi}_T}+K_{\rho}q^2\right)\widetilde{\delta \rho}
        +
        \beta_V\widetilde{\delta T}
        -   \widetilde{\sigma}_{\parallel, u}
    \right],
\end{eqnarray}
where \add{in the last step we take the long-wavelength limit} $q\ll 1$, since we are concerned only with the large-scale behavior of the system.
The inverse Fourier transform of Eq.~(\ref{delta_phi}) yields Eq.~(\ref{RCH-AA}), where $\mu_b=\frac{\delta f_b}{\delta \rho}$ is the effective chemical potential with $f_b$ the effective bulk free energy density
\begin{eqnarray}
    f_b= \frac{\sigma^2}{2\bar{\chi}_T}\rho^2 + \int \sigma^2\beta_V T_t(\rho)d\rho.
\end{eqnarray}

\section*{Appendix D: Universality of Absorbing Phase Transition}

In the simulations of spinner fluids, we obtain critical exponents of the absorbing transition, which is consistent with the CDP universality class ({Fig.~S3 in SM~\cite{supmat}})~\cite{henkel2008non}. {Here, we give the theoretical explanation.}
As shown in Fig.~\ref{MainFigHydro}(b), near the critical point,  we have $\delta T_t/\delta \rho \gg 1$. Therefore, the $\delta\rho$ term in the right side of Eq.~(\ref{delta_phi}) can be ignored, which gives
\begin{eqnarray}
    -i\omega \delta \rho 
    \approx 
    -\frac{q^2}{\gamma_t} 
    \left(
        \beta_V\widetilde{\delta T}
        -   \widetilde{\sigma}_{\parallel, u}
    \right).
\end{eqnarray}
The inverse Fourier transform of the above equation yields
\begin{eqnarray} \label{TDiffu}
    \frac{\partial  \rho}{\partial t}
    =    M\nabla^2T_t+\Gamma\nabla^2 \sigma_{\parallel, u}
\end{eqnarray}
with $M= {\beta_V}/{\gamma_t}$ and $\Gamma = -  \gamma_t^{-1}$.  As shown later, for the absorbing transition, $\Gamma\nabla^2 \sigma_{\parallel, u}$ is a higher order noise term and can be neglected. \add{Since the system
enters the absorbing state when $T_t=0$}, the noise in Eq.~(\ref{kineticEqn2}) must be multiplicative, i.e., $\langle \eta_t(\mathbf{x},t)\eta_t(\mathbf{x}',t') \rangle=
    F_t(T_t)\delta^d(\mathbf{x}'-\mathbf{x})\delta(t-t')$ with $F_t(0)=0$.
Neglecting the effect of convection and expanding $T_r(\rho,T_t)$ and $\Bar{Z}(\rho,T_t)$ near the critical point $\rho={\rho}_c,~T_t=0$, we obtain
\begin{eqnarray}
    \bar{Z} &=& z_1 T_t + z_2(\rho-\rho_c)+\cdots,\\
         T_r &=& T_{ss} + t_1T_t + t_2(\rho-\rho_c)+\cdots
\end{eqnarray}
\add{Substituting above expanding into Eq.~(\ref{kineticEqn1}) and} ignoring higher-order terms, we can get
\begin{align}\label{CDP_1}
    \frac{\partial T_t}{\partial t} =& D_t\nabla^2 T_t - u_0  - rT_t - u_1T_t^2 + u_2 \rho T_t \nonumber\\
    &+ 
    u_3\rho + u_4\rho^2 +
    \eta_t(\mathbf{x},t).
\end{align}
One can notice that $r=2\gamma_t/m-s_3 z_1 T_{ss}$, so the larger the friction $\gamma_t$, the easier it is for the system to fall into the absorbing state; similarly, the larger the driving power $T_{ss}$, the easier it is for the system to enter the active state, consistent with the simulation results. 
\add{Since} the spinner system falls into the absorbing state when $T_t=0$ regardless of the density distribution, $\partial_t T_t=0$ must be vanish at $T_t=0$ for arbitrary $\rho$, i.e., $-u_0+u_3\rho+u_4\rho^2=0$.
Combining Eq.~(\ref{TDiffu}) and Eq.~(\ref{CDP_1}), the field equations for the system near the absorbing critical point can be written as
\begin{align}
    &\frac{\partial  \rho}{\partial t}
    =
    M\nabla^2T_t\label{CDP1}\\
    &\frac{\partial T_t}{\partial t} = D_t\nabla^2 T_t - rT_t - u_1T_t^2 + u_2 \rho T_t + \eta_t(\mathbf{x},t)\label{CDP2}\\
    &\langle \eta_t(\mathbf{x},t)\eta_t(\mathbf{x}',t') \rangle
    =
    \sigma T_t\delta^d(\mathbf{x}'-\mathbf{x})\delta(t-t').  \label{CDP3}
\end{align}
Note that in the above, we only keep the lowest order term in noise terms, because higher order terms are irrelevant in the renormalization-group sense. 
Eqs.~(\ref{CDP1}-\ref{CDP3}) are essentially equivalent to \add{field equations} of the Reggeon field theory~\cite{PhysRevE.62.R5875}, suggesting that the absorbing transition of active spinner fluids belongs to the CDP universality class.

\section*{Appendix E: Dynamical Field Theory and Renormalization-Group Method}

\subsection{Dynamical Field Theory for Generalized Model B with Center-of-Mass Conserved Noise}

As an extension \add{of the formalism used for} classical Model B in Ref.~\cite{CriDyn}, we now introduce the dynamical field theory of the generalized Model B 
\begin{eqnarray}\label{CMC-modelB-AP}
    \frac{\partial \psi}{\partial t}
    =
    \nabla^2\left({
        r\psi - \nabla^2 \psi + \frac{u}{6} \psi^3 - h
    }\right)
    + 
    {\zeta}(\mathbf{x},t).
\end{eqnarray}
{For simplicity, here we set the dimensionless \add{prefactor of the gradient term} to $\kappa=1$ in subsequent theoretical analysis, which means that \add{we} rescale time unit $\Tilde{\tau}_0=\sigma^4$. }$h\to 0$ is a weak external field and noise correlation is
\begin{eqnarray}\label{CMC-modelB-AP-noise}
    \langle{ \zeta(\mathbf{x},t)\zeta(\mathbf{x}',t') }\rangle = 2D (i\nabla)^{2+\theta}\delta^d(\mathbf{x}-\mathbf{x}')\delta(t-t').
\end{eqnarray}
The imaginary unit $i$  is introduced only for simplification of the notation in momentum space~\cite{hohenberg1977theory,CriDyn}. Here, $\theta=0$ corresponds to the classical Model B~\cite{hohenberg1977theory,CriDyn}, while $\theta=2$ corresponds to Eq.~(\ref{CMC-modelB-s}), the proposed Model B-like theory for HU fluids.

To facilitate the renormalization group computations below, we now introduce the field-theoretic formalism based on Martin-Siggia-Rose-Jansen-De Dominicis (MSRJD) action~\cite{CriDyn}. 
The main idea is  to write the probability distribution of the random variable field $\psi$ described by Eq.~(\ref{CMC-modelB-AP}) as an integration of the auxiliary field $\Tilde{\psi}$ with statistical weight determined by MSRJD action $\mathcal{A}[\Tilde{\psi},\psi]$, i.e., 
\begin{eqnarray}
    \mathcal{P}_{\psi}[\psi]
    =
    \mathcal{C}^{-1}\int \mathcal{D}[i\Tilde{\psi}]e^{-\mathcal{A}[\Tilde{\psi},\psi]}.
\end{eqnarray}
The MSRJD action can be divided into Gaussian (linear) and non-Gaussian (nonlinear) parts $\mathcal{A}[\Tilde{\psi},\psi]=\mathcal{A}_0[\Tilde{\psi},\psi]+\mathcal{A}_{I}[\Tilde{\psi},\psi]$. The Gaussian part is
\begin{eqnarray}\label{action_0}
\begin{aligned}
        \mathcal{A}_0[\Tilde{\psi},\psi]
    =&
    \int d^dx\int dt\ \left(
        \Tilde{\psi}(\mathbf{x},t)\left[{
            \frac{\partial }{\partial t}-\nabla^2(r-\nabla^2)    
        }\right]  \psi(\mathbf{x},t) \right.\\
        &\left.-
        D\Tilde{\psi}(\mathbf{x},t)(i\nabla)^{2+\theta}\Tilde{\psi}(\mathbf{x},t)
        +
        \Tilde{\psi}\nabla^2 h(\mathbf{x},t) \right).
\end{aligned}
\end{eqnarray}
The non-Gaussian part is
\begin{eqnarray}\label{action_ini}
    \mathcal{A}_{I}[\Tilde{\psi},\psi]
    =
    -\frac{u}{6}\int d^dx\int dt\
    \Tilde{\psi}(\mathbf{x},t) { \nabla^{2} } [{\psi}(\mathbf{x},t)]^3.
\end{eqnarray}
We first consider the Gaussian field theory, in which the bare response propagator and bare correlation function are defined as
\begin{align}
    &\langle{{\psi}(\mathbf{q},\omega)\Tilde{\psi}(\mathbf{q}',\omega')}\rangle_0
    =
    G_0(q,\omega)(2\pi)^{d+1}\delta^d(\mathbf{q}+\mathbf{q}')\delta(\omega+\omega')
    \\
    &\langle{{\psi}(\mathbf{q},\omega){\psi}(\mathbf{q}',\omega')}\rangle_0
    =
    C_0(q,\omega)(2\pi)^{d+1}\delta^d(\mathbf{q}+\mathbf{q}')\delta(\omega+\omega')
\end{align}
which yield the bare response propagator and bare correlation function
\begin{eqnarray}
    G_0(q,\omega)
    &=&
    \frac{1}{ -i\omega + q^2(r+q^2) }\label{Sus0q} \\
    C_0(q,\omega)
    &=&
    \frac{2D q^{2+\theta}}{ \omega^2 + [ q^2(r+q^2) ]^2 } \nonumber\\&=& 2Dq^{2+\theta} G_0(q,\omega)G_0(-q,-\omega).  \label{C0q}
\end{eqnarray}
In the field theory, perturbations can be represented as Feynman diagrams.
Here we introduce elements of Feynman diagrams of our field theory:
\begin{eqnarray}\label{FeynRule}
    \includegraphics[width=85mm]{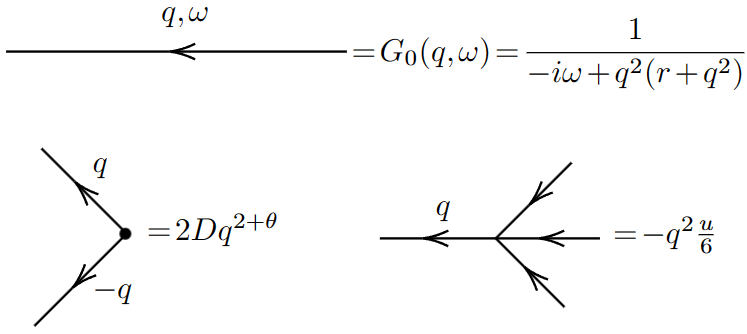}
\end{eqnarray}
Compared with the $O(1)$-symmetric relaxational Model B in Ref.~\cite{CriDyn}, our model differs only in that the two-point noise vertex is multiplied by a $q^{\theta}$ factor.

\subsection{Response Functions and Correlation Functions}

The dynamical susceptibility in real space is defined as
\begin{eqnarray}\label{DefinitionSus}
    \chi(\mathbf{x}-\mathbf{x}',t-t') = \left.\frac{\delta\langle \psi(\mathbf{x},t) \rangle}{\delta h(\mathbf{x}',t')}\right|_{h=0}.
\end{eqnarray}
In Fourier space, the bare susceptibility is
\begin{eqnarray}\label{suscep0}
\chi_0(\mathbf{q},\omega)  =  q^2G_0(q,\omega).
\end{eqnarray}
The {static} bare susceptibility function  can be obtained by the inverse Fourier transformation
\begin{eqnarray}\label{sus}
    \chi_0(x)
    &=&
    \int\frac{d^dq}{(2\pi)^d} e^{-i\mathbf{q}\cdot\mathbf{x}}   \chi_0(q,\omega=0) \nonumber\\
    &=&\int_q \frac{e^{-i\mathbf{q}\cdot\mathbf{x}}}{\xi^{-2}+q^2}  
    \sim 
    \left\{
    \begin{aligned}
        \dfrac{1}{x^{d-2}},\ \ \ 
    x\ll \xi \\
    \dfrac{e^{-x/\xi}}{x^{(d-1)/2}},\ \ \ 
    x\gg \xi \\
    \end{aligned}
    \right.
\end{eqnarray}
where the characteristic length $\xi=r^{-1/2}$ is the correlation length. Here we use the asymptotic behavior of the Ornstein-Zernike correlation function.  {The bare static structure factor is
\begin{equation}\label{bareStaticStructureFactor}
    S_0(q)=\int\frac{d\omega}{2\pi} C_0(q,\omega)=\int\frac{d\omega}{2\pi} \frac{2D q^{2+\theta}}{ \omega^2 + [ q^2(r+q^2) ]^2 }
    =
    \frac{Dq^{\theta}}{r+q^2}.
\end{equation} }
When $\theta=0$, the bare correlation function in real space can be obtained asymptotically
\begin{eqnarray}
    C_0(x)
    &=&
    \int\frac{d^dq}{(2\pi)^d} e^{-i\mathbf{q}\cdot\mathbf{x}} \int\frac{d\omega}{2\pi}  C_0(q,\omega) \nonumber\\
    &=&D\int_q \frac{e^{-i\mathbf{q}\cdot\mathbf{x}}}{r+q^2} 
    \sim
    \left\{
    \begin{aligned}
        \dfrac{D}{x^{d-2}},\ \ \ 
    x\ll \xi \\
    \dfrac{De^{-x/\xi}}{x^{(d-1)/2}},\ \ \ 
    x\gg \xi \\
    \end{aligned}
    \right.
\end{eqnarray}
For $\theta=2$ we have
\begin{eqnarray}
    C_0(x)
    &&=  D\int_q \frac{q^2e^{-i\mathbf{q}\cdot\mathbf{x}}}{r+q^2} 
    =  D\int_q e^{-i\mathbf{q}\cdot\mathbf{x}}    
    - 
    Dr\int_q \frac{e^{-i\mathbf{q}\cdot\mathbf{x}}}{r+q^2}\nonumber\\  
    &&\sim
    \left\{
    \begin{aligned}
        &D\delta(\mathbf{x})-\dfrac{Dr}{x^{d-2}},\ \ \ 
    & x\ll \xi \\
    &D\delta(\mathbf{x})-\dfrac{Dre^{-x/\xi}}{x^{(d-1)/2}},\ \ \ 
    & x\gg \xi \\
    \end{aligned}
    \right. \label{C0x}
\end{eqnarray}
At the critical point, any scale must be washed away, which indicates that the correlation function takes the form of a homogeneous function. This scaling hypothesis holds for both $\theta=0$ and $\theta=2$
\begin{eqnarray}\label{ScalingForm}
    C(x,r)=\frac{\hat{\mathbf{C}}(xr^{\nu})}{x^{d+\theta-2+\eta}}.
\end{eqnarray} 
For the case of $\theta=2$ and $d\geqslant2$ (mean-field), there is $\nu=1/2, ~\eta=0$, and the correlation function of the mean-field Eq.~(\ref{C0x}) also satisfies the form Eq.~(\ref{ScalingForm})
\begin{eqnarray}
    C(x,r)=\frac{\hat{\mathbf{C}}(xr^{1/2})}{x^{d}}
    \sim
    \delta^d(\mathbf{x})-\frac{Dr}{x^{d-2}},\ \ \ 
    x\ll \xi
\end{eqnarray}
i.e.,
\begin{align}
    \hat{\mathbf{C}}(xr^{1/2}) &\sim 
    x^d\delta^d(\mathbf{x}) - \frac{Dr}{x^{-2}}\nonumber\\
    &=
    (xr^{1/2})^d\delta^d(\mathbf{x}r^{1/2})-\frac{D}{(xr^{1/2})^{-2}}
    ,\ \ \ 
    x\ll \xi
\end{align}
which means
\begin{eqnarray}\label{HomoForm}
    \hat{\mathbf{C}}(X) \sim X^d\delta^d(\mathbf{X})-\frac{D}{X^{-2}},\ \ \ 
    X\ll \xi
\end{eqnarray}
For the general form when $\theta=2$, inserting Eq.~(\ref{HomoForm}) into Eq.~(\ref{ScalingForm}) gives
\begin{eqnarray}
    C(x,r) = \frac{\hat{\mathbf{C}}(xr^{\nu})}{x^{d+\eta}}
    \sim
    \frac{\delta^d(\mathbf{x})}{x^{\eta}} - \frac{Dr^{2\nu}}{x^{d-2+\eta}},\ \ \ 
    x\ll \xi.
\end{eqnarray}
{For $\theta=0$, the classical form $C(x)\sim x^{-d+2-\eta}$ can be obtained by the same manner.}

\subsection{Dimensional Analysis}

In this section, we perform a dimensional analysis of the field theory under the scaling transformation $\mathbf{x} \to  b\mathbf{x'} $. We denote the scaling dimension of $v$ by $\Delta_v$, with $\Delta_x=-\Delta_q=-1$. The invariance of the theory requires the MSRJD action Eq.~(\ref{action_0}) to remain invariant, \add{implying}  $\Delta_{\mathcal{A}}=0$~\cite{CriDyn}. The invariance of the MSRJD action further requires the invariance of each term $\int d^dx\int dt~\Tilde{\psi}\partial_t\psi,~\int d^dx\int dt~\Tilde{\psi}\nabla^4\psi,~\int d^dx\int dt~D\Tilde{\psi}(i\nabla)^{2+\theta}\Tilde{\psi}$ in Eq.~(\ref{action_0}). Without considering anomalous dimension, this gives
\begin{align}
     &-d+\Delta_t+\Delta_{\Tilde{\psi}}-\Delta_t+\Delta_{\psi} = 0  \Rightarrow  \Delta_{\Tilde{\psi}}+\Delta_{\psi}=d\\
     &-d+\Delta_t+\Delta_{\Tilde{\psi}}+\Delta_{\psi}+4= 0  \Rightarrow  \Delta_t=-\Delta_{\omega}=-4\\
     &-d+\Delta_t+\Delta_D+2\Delta_{\Tilde{\psi}}+2+\theta = 0 \nonumber\\  &\Rightarrow  \Delta_{\Tilde{\psi}}=\frac{d+2-\theta}{2},\Delta_{\psi}=\frac{d-2+\theta}{2}.
\end{align}
Note that $\Delta_D=0$, since the noise strength $D$ should be invariant under the scaling transformation~\cite{CriDyn}. Next we discuss the finite-size scaling of density
fluctuations, or equivalently, static structure factor
\begin{align}\label{FSS-Sq}
    S(\tau,q)
    &=
    \int \frac{d\omega}{2\pi} C(q,\omega)
    =
    \int \frac{d\omega}{2\pi}
    \int d^dx\int dt~e^{-i\boldsymbol{q}\cdot\boldsymbol{x} -i\omega t}
    C(x,t)\nonumber\\
    &=
    \int \frac{d\omega}{2\pi}
    \int d^dx\int dt~e^{-i\boldsymbol{q}\cdot\boldsymbol{x} -i\omega t}
    \langle \psi(x,t)\psi(x',t') \rangle\ .
\end{align}
We can obtain the scaling dimension of the static structure factor  $\Delta_S=
    \Delta_{\omega}+\Delta_{C}=\Delta_{\omega}-d+\Delta_t+2\Delta_{\psi}=\theta-2$. 
It should be noted that below the critical dimension, fluctuations cause the dimension of the actual correlation function to deviate from that predicted by mean-field theory, i.e.,
\begin{eqnarray}\label{CqEta}
    C(q,\omega)\sim q^{\Delta_{C}+\eta}
\end{eqnarray}
where $\eta$ is the anomalous dimension reflecting the difference between the actual correlation function and the mean-field result. According to Eqs.~(\ref{FSS-Sq}-\ref{CqEta}), we have
\begin{eqnarray}
    S(\tau,q)\sim q^{\Delta_S+\eta}.
\end{eqnarray}
Thus, the static structure factor has the homogeneous form
\begin{eqnarray}\label{PrimeScaling}
    S(\tau,q) 
    = 
    q^{\theta+\eta-2}\mathcal{S}(\tau/q^{1/\nu})
    =
    q^{\theta-\gamma/\nu}\mathcal{S}(\tau/q^{1/\nu})
\end{eqnarray}
where we use the scaling relationship $\gamma=\nu(2-\eta)$. The  amplitude of the static structure factor at the minimum wave number $2\pi/L$, i.e., $\chi_{\rho}(\tau,L)=S(\tau,2\pi/L)$, satisfies~\cite{Torquato2003}   
\begin{eqnarray}
    \chi_{\rho}(\tau,L) 
    =
    L^{\gamma/\nu-\theta}{S}_{\rho}(\tau L^{1/\nu}).
\end{eqnarray}
Finally, we address whether the addition of active flux terms $a_1\nabla^2\left( \nabla \psi \right)^2 - a_2 \nabla\cdot \left[ (\nabla^2\psi)\nabla\psi \right]$ ~\cite{wittkowski2014scalar,cates2025active} to the MSRJD action Eq.~(\ref{action_0}) affects the LG criticality of HU fluids. Taking $a_1\nabla^2\left( \nabla \psi \right)^2$ as an example, this term is written as $\int d^dx\int dt~a_1\Tilde{\psi}\nabla^2\left( \nabla \psi \right)^2$ in the  MSRJD action. From the dimensional analysis \add{above, we have} $\Delta_{\psi}=\Delta_{\Tilde{\psi}}=d/2$ for HU fluids ($\theta=2$), \add{which gives} $-d-4+\Delta_{a_1}+\Delta_{\Tilde{\psi}}+4+2\Delta_{\psi}=0$. So $\Delta_{a_1}=-d/2<0$, which means that this term is irrelevant for LG criticality. The same method also yields $\Delta_{a_2}=-d/2<0$. This implies that active flux terms do not affect the LG criticality of HU fluids. 

In general, energy fluctuations obey the finite-size scaling $c_v(\tau,L) = f(\tau) + L^{-\lambda}g(\tau L^{1/\nu})$~\cite{kimball2005universality}. Expanding the function $g(\tau L^{1/\nu})$ gives $c_v(\tau,L) = f(\tau) + c_0 L^{-\lambda} + \sum^{\infty}_{i=1} c_i \tau^i L^{i/\nu-\lambda}$.  
Since $c_v(T,L)$ does not diverge with $L$ in Fig.~5(h), high order terms with $i/\nu - \lambda >0$ should be neglected. By fitting the data \add{to} $c_v(\tau, L) = f(\tau) + c_0 L^{-\lambda} + c_1 \tau L^{1/\nu - \lambda}$, we determine that $\lambda=2.11(2),~c_0=24.0(5),~c_1=13.0(2)$.

\subsection{Wilsonian Momentum Shell Renormalization-Group Method}

We \add{begin by calculating} the flow equations of the couplings using Wilsonian momentum shell renormalization-group (RG) approach~\cite{wilson1983renormalization,Papanikolaou2023}.
The first step of RG is coarse-graining, i.e., integrating out the wave vectors in an infinitesimal shell $q \in [\Lambda/b,\Lambda]$ where $b=e^{\delta l}=1+\delta l$ and  $\Lambda$ is the ultraviolet cutoff of the system.  Let $v$ be a coupling in the action ($v=$ $r$ or $u$), then its linearly corrected coupling $\Tilde{v}$ due to the infinitesimal coarse-graining is $\Tilde{v}=v+C_{v}v\delta l$
with $C_{v}$ quantifying graphical one-loop corrections for $v$. 
Here we only need to consider one-loop graphs, since multi-loops graphs are higher order corrections and can be neglected. The next step is to restore the ultraviolet cutoff $\Lambda/b \to \Lambda$ through rescaling
\begin{align}
    v'&=b^{\Delta_{v}}\Tilde{v}
    =
    (1+\delta l)^{\Delta_{v}}(1+C_{v}\delta l)v\nonumber\\
    &=
    [1+(\Delta_{v}+C_{v})\delta l]v + O(\delta l^2).
\end{align}
Then we can get the Wilsonian flow equation
\begin{eqnarray}\label{DefineFlowEquations}
    \partial_l v = (\Delta_{v}+C_{v})v.
\end{eqnarray}
This equation describes the change of couplings like $r$ or $u$ under the renormalization as $l$ increases. We introduce the reduced couplings and fields to make the Wilsonian flow equations more concise
\begin{align}
    &\bar{\Tilde{\psi}}=\Tilde{\psi}{D^{1/2}}\Lambda^{-(d-\theta+2)/2},~\bar{\psi}=\psi {D^{-1/2}}\Lambda^{-(d+\theta-2)/2}\nonumber\\
    &\overline{r}=\frac{r}{\Lambda^2},~
    \overline{u}=u {D K_d} \Lambda^{-\varepsilon}
\end{align}
where $\varepsilon = d_c  -d$, $ d_c =4 - \theta$ is the upper critical dimension and $\Lambda$ is the ultraviolet cutoff. Note that the flow equations are independent of $\theta$, whose sole effect under the renormalization is to change $d_c$. Based on one-loop perturbations for the propagator and the four-point vertices, the flow equations can be obtained as~\cite{CriDyn}
\begin{eqnarray}\label{flowEquations}
     \partial_l \overline{r} = 2\overline{r}+\frac{\overline{u}}{2}\frac{1}{\overline{r}+1},\ \ \ 
     \partial_l\overline{u}
     =
     \left(
     {
         \varepsilon - \frac{3\overline{u}}{2}
         \frac{1}{(\overline{r}+1)^2}
     }\right)
     \overline{u}
\end{eqnarray}
When $d> d_c$, i.e., $\varepsilon<0$, the Gaussian fixed point $G$ ($\overline{u}^*_G=\overline{r}^*_G=0$) is the only IR stable fixed point, which corresponds to the mean-field theory. When $d < d_c$, the IR stable fixed point is $P_{\theta}$ ($\overline{u}^*_{\theta}=2\varepsilon/3$, $\overline{r}^*_{\theta}=-\varepsilon/6$), and when $\theta=0$ it is just the Wilson-Fisher (WF) fixed point~\cite{CriDyn}.   
Linearized flow equations are
\begin{eqnarray}
    \partial_l\begin{pmatrix}
        \delta\bar{r} \\
        \delta\bar{u}
    \end{pmatrix}
    =
    \begin{pmatrix}
        2-\dfrac{\varepsilon}{3} &\dfrac{1}{2}\left({ 1+\dfrac{\varepsilon}{6} }\right) \\
        0 &-\varepsilon
    \end{pmatrix}
    \begin{pmatrix}
        \delta\bar{r} \\
        \delta\bar{u}
    \end{pmatrix}.
\end{eqnarray}
These equations have two eigenvalues
\begin{eqnarray}
    \Delta_{\delta \bar{u}}=-\varepsilon,\ \ \ 
    \Delta_{\delta\bar{r}} = 2-\frac{\varepsilon}{3}.
\end{eqnarray}
We aim to calculate critical exponents associated with the correlation length, order parameter , susceptibility, and field response
\begin{eqnarray}
    \xi\sim\tau^{-\nu},~\psi\sim\tau^{\beta},~\chi_c\sim\tau^{-\gamma},~\psi\sim h^{1/\delta}
\end{eqnarray}
where $\tau=|r-r_c|$. Using the results of the scaling dimensional analysis, we can obtain the scaling relations
\begin{align}
        &\gamma=\beta(\delta-1),~~
    \gamma=\nu(2-\eta),\nonumber\\
    &2\beta=(d-2+\theta+\eta)\nu,~~ z=4-\eta. \label{scalingRelation}
\end{align}
From Eq.~(\ref{scalingRelation}),  we obtain critical exponents to lowest order~\cite{CriDyn},
\begin{equation}
    \begin{aligned}
        &\nu = \frac{1}{2}+\frac{\varepsilon}{12}+O(\varepsilon^2), ~~ 
    \eta=\frac{\varepsilon^2}{54}+O(\varepsilon^3), \\
    &\beta  = \frac{1}{2}-\frac{\varepsilon}{6}+O(\varepsilon^2),~~
    \gamma =1+\frac{\varepsilon}{6}+O(\varepsilon^2), \\
    &\delta = 3+\varepsilon+O(\varepsilon^2)~~,
    z= 4+O(\varepsilon^2)
    \end{aligned}
\end{equation}

\subsection{The Logarithmic Correction to the 2D HU Fluids Correlation Function}

\add{Different from equilibrium cases~\cite{kenna2006self}}, logarithmic corrections to the mean-field correlation function Eq.~(\ref{MeanFieldCorrelation}) arise at the upper critical dimension $d=2$ due to corrections of $r$ for HU fluids.
Here we use the renormalization-group method in Ref.~\cite{toner2023roughening,chen2025order} to estimate these corrections.
For $d=2$, the renormalization-group flow equations Eqs.~(\ref{flowEquations}) near the fixed point can be expanded as
\begin{align}
     &\partial_l \delta\bar{r}(l) = 
     2\delta\bar{r}(l) + \frac{\delta\bar{u}(l)}{2} - \frac{1}{2}\delta\bar{r}(l)\delta\bar{u}(l)\nonumber\\
     &+O(\delta\bar{u}(l)\delta\bar{r}(l)^2,\delta\bar{u}(l)^2)\label{LogRG1_0}  \\
     &\partial_l\delta\bar{u}(l)
     =
     -\frac{3}{2}\delta\bar{u}(l)^2 +O(\delta\bar{u}(l)^2y(l)). \label{LogRG2}
\end{align}
Performing the linear transformation $y(l)=\delta\bar{r}(l)+\delta\bar{u}(l)/4$ to Eq.~(\ref{LogRG1_0}), we obtain
\begin{eqnarray}
     \partial_l y(l) = \left(2-\frac{1}{2}\delta\bar{u}(l)  \right)y(l)+O(\delta\bar{u}(l)y(l)^2,\delta\bar{u}(l)^2).~~~~~~~ \label{LogRG1}
\end{eqnarray} 
Solving Eq.~(\ref{LogRG2}) yields
\begin{eqnarray}\label{gl}
    \delta\bar{u}(l)=\left.\bar{u}(l)-\bar{u}^*\right|_{\varepsilon=0}=\bar{u}(l)=\frac{\delta\bar{u}_0}{\frac{3}{2}\delta\bar{u}_0l+1}.
\end{eqnarray}
where {$\delta\bar{u}_0=\delta\bar{u}(l=0)=\frac{D \delta{u} }{2\pi}$} is the bare value (the parameter of the actual system).
Inserting Eq.~(\ref{gl}) into Eq.~(\ref{LogRG1}) yields
\begin{eqnarray}
    \frac{dy(l)}{y(l)}
    =
    \left( 2 - \frac{\delta\bar{u}_0}{3\delta\bar{u}_0l+2} \right) dl.
\end{eqnarray}
Solving the above equation yields
\begin{eqnarray}
\delta\bar{r}(l)
=
\delta\bar{r}_0 \frac{ e^{2l} }{ \left[ \frac{3}{2}\delta\bar{u}_0 l + 1 \right]^{1/3} } 
- \frac{\delta\bar{u}(l)}{4}.
\end{eqnarray}
Multiplying both sides by $\Lambda^2$ gives $\delta\bar{r}(l)\to\delta{r}(l)$. 
\add{Since} Eq.~(\ref{gl}) indicates that the nonlinear $u$ term decays to zero under the RG flow ($l\to\infty$), we have
\begin{eqnarray}\label{rl}
\delta{r}(l) 
=\left.\bar{r}(l)-\bar{r}^*\right|_{\varepsilon=0}
=
{r}(l)
=
\delta{r}_0 \frac{ e^{2l} }{ \left[ \frac{3}{2}\delta\overline{u}_0 l + 1 \right]^{1/3} }~~~
\end{eqnarray}
where $\delta{r}_0=\delta{r}(l=0)$. Near the critical point ($\delta r_0\ll 1,~\xi\gg1$), let $l_c=\ln \xi\mu$, where $\mu$ is an arbitrary momentum scale~\cite{CriDyn}. Then we can obtain
\begin{eqnarray}
    \delta{r}(l_c) \mu^{-2} = \delta{r}_0 \xi ^2 \left( \frac{3}{2}\delta\bar{u}_0 \ln \xi\mu \right)^{-1/3},
\end{eqnarray}
This gives
\begin{eqnarray}
    \delta{r}_0 \sim \xi^{-2} ( \ln \xi )^{1/3}.
\end{eqnarray} 
Upon iteratively solving for the correlation length~\cite{CriDyn}, we have the corrected correlation length
\begin{eqnarray}
    \xi \sim \delta{r}_0^{-1/2}| \ln \delta{r}_0 |^{1/6}.
\end{eqnarray}
We further express the correlation function at different steps of the RG~\cite{nelson1975crossover}
\begin{eqnarray}\label{RGflowCorr}
    C(x,\delta{r},u)
    =
    e^{-2l}C(e^{-l}x,\delta{r}(l),u(l))
\end{eqnarray}
This relation should be exact at a fixed point (otherwise, the scaling invariance will be violated).
The RG parameter $l$ can be chosen arbitrarily, and $e^l=x\Lambda$ is a convenient choice~\cite{toner2023roughening}. According to Eq.~(\ref{gl}), this choice ensures that $l$ is always large enough that the nonlinear $u$ term can be neglected.
This in turn implies that the correlation function on the right-hand side of Eq.~(\ref{RGflowCorr}) can be evaluated using the mean-field result Eq.~(\ref{C0x}), i.e.,
\begin{align}
    C(x,\delta{r},u)
    &\sim
    De^{-2l}\delta^2(e^{-l} \mathbf{x})
    -
    e^{-2l} D\delta{r}_0  
    \frac{ e^{2l} }{ \left[ \frac{3}{2}\delta\overline{u}_0 l + 1 \right]^{1/3} } \nonumber\\
    &\sim
    D\delta^2(\mathbf{x})
    -
    \frac{D\delta{r}_0 }{\left[  \add{\frac{3}{2}\delta\overline{u}_0} \ln(\Lambda x) \right]^{1/3}},\ \ \ 
     x\ll \xi. \label{Cxlog}
\end{align}
In the last equation, we used $\frac{3}{2}\delta\overline{u}_0 l + 1\sim  \add{\frac{3}{2}\delta\overline{u}_0}l~(l\to\infty)$.

\bibliographystyle{nature}
\bibliography{reference}

\end{document}